\documentstyle[prd,aps,epsfig,psfig]{revtex}
\begin{document} 
\draft 
\title{Photon mixing in universes with large extra--dimensions}   
\author{C\'edric  
Deffayet$^1$ and Jean--Philippe Uzan$^{1,2}$}   
\address{(1) Laboratoire  
de Physique Th\'eorique\footnote{Unit\'e Mixte de Recherche du CNRS, UMR 
8627.}, 
 Universit\'e Paris XI, B\^at. 210 \\ F--91405  
Orsay Cedex (France).\\ (2) D\'epartement de Physique Th\'eorique,  
Universit\'e de Gen\`eve,\\ 24 quai E. Ansermet, CH--1211 Gen\`eve 4  
(Switzerland).}    
\date{\today}   
\maketitle  
 
\begin{abstract}  
In presence of a magnetic field, photons can mix with any particle
having a two--photon vertex. In theories with large compact
extra--dimensions, there exists a hierachy of massive Kaluza--Klein
gravitons that couple to any photon entering a magnetic field. We study
this mixing and show that, in comparison with the four dimensional
situation where the photon couples only to the massless graviton, the
oscillation effect may be enhanced due to the existence of a large
number of Kaluza--Klein modes.  We give the conditions for such an
enhancement and then investigate the cosmological and astrophysical
consequences of this phenomenon; we also discuss some laboratory
experiments.  Axions also couple to photons in the same way; we discuss
the effect of the existence of bulk axions in universes with large
extra--dimensions. The results can also be applied to neutrino physics
with extra--dimensions.
\end{abstract}  
\pacs{ 
      {\bf Preprint numbers:} UGVA--DPT 99/10-1053, LPT--ORSAY 99/104} 
\vskip2pc

\section{Introduction}  
  
It is well known \cite{gertsenshtein62,zeldovich83} that photons can be 
converted into gravitons by a magnetic field in a standard four 
dimensional spacetime. The propagation eigenstates are then mixtures of 
photon and graviton interaction eigenstates. In quantum words, this 
mixing is due to the fact that any particle which has a two--photon 
vertex can be created by a photon entering an external electromagnetic 
field \cite{adler71,raffelt88} and can oscillate coherently with the 
photon. Classically, this can be understood by the fact that an 
electromagnetic plane wave cannot radiate gravitationally in vacuum 
since its stress--energy tensor contains no quadrupole 
\cite{landau}. But, a time varying quadrupole appears (due to 
interference) when an electromagnetic plane wave propagates through a 
constant magnetic field \cite{gertsenshtein62,zeldovich83,magueijo94}. 
The implications of this effect on the cosmic microwave background 
(photons) has been considered \cite{magueijo94,chen95} and it has been 
shown that it will be undetectable for standard cosmological magnetic 
fields \cite{cillis96}. A similar effect also generically happens for 
axions (and for any particle having a two--photon vertex).  The 
photon--axion (see \cite{kim87,raffelt90,turner90} for reviews on 
axions) mixing has yet been studied in details by many authors (see 
e.g. \cite{raffelt88,sikivie83,sikivie84,morris}) and is used in 
experiments, since the pioneer work by Sikivie \cite{sikivie83}, to put 
constraints on the axion parameters 
\cite{bibber89,hagmann98,kim98,sikivie97} (see e.g. \cite{raffelt98} for 
an up to date review on such experiments). 
 
 Recently, a lot of interest has been raised by models where the 
universe has large extra--dimensions \cite{arkani98,arkani99}.  In such 
models, the Planck scale, $M_4$, is no longer a fundamental scale but is 
related to the fundamental mass scale of the $D$ dimensional theory, 
$M_D$, through \cite{arkani99} 
\begin{equation}\label{planck}  
\bar M_4^2\equiv R^n M_D^{n+2},  
\end{equation}  
where $R$ is a length scale (usually taken to be the radius of the 
$n=D-4$ compact extra--dimensions) and $\bar M_4\equiv 
M_4/\sqrt{8\pi}$. $M_D$ can be significantly smaller than $\bar M_4$ at 
the price of having large extra--dimensions. These ideas can be 
naturally embedded in fundamental string theories with a low string 
scale \cite{witten96,lykken96,dienes98,antoniadis98} (see also 
\cite{antoniadis90} for earlier discussions on TeV scale 
extra--dimensions).  In these models, gravity can propagate in the $D$ 
dimensional spacetime (bulk space time) whereas the standard model 
fields are localised on a 3--brane.  An effect of the compact 
extra--dimensions arises from interactions between the Kaluza--Klein 
(KK) excitations of the gravitons (or other bulk fields) which are seen 
in four dimensions as a tower of massive particles 
\cite{giudice98,han98}.  Constraints on the size of these 
extra--dimensions can be obtained both from the laboratory physics 
\cite{arkani99} and from astrophysics and cosmology (see 
e.g. \cite{bdl}). For instance, the emission of KK gravitons induces an 
energy loss in many astrophysical objects \cite{barger99} such as the 
Sun, red giants and supernovae SN1987A \cite{cullen99} implying the 
lower bound $M_D>30-130\,\hbox{Tev}$ (2.1 - 9.2) TeV for the case of 
$n=2$ (3) extra--dimensions \cite{barger99,cullen99}. Some authors 
\cite{arkani98,antoniadis98,chang99,dudas2} also have recently 
investigated the possible presence of axions in the bulk which would be 
coupled to the brane degrees of freedom. Such a bulk axion also gives 
rise to a tower of KK-states as seen from a four dimensional point of 
view. 
  
The goal of this article is to investigate the effects of the photon--KK 
graviton and photon--KK axion oscillations and to estimate their 
effects in cosmology and astrophysics, as well as terrestrial 
experiments. Since there is a large number of KK states with which the 
photon can mix, one can expect a departure from the usual four 
dimensional result. We first describe (\S~\ref{par2}) the photon--KK 
graviton system starting from a $D$ dimensional action, linearising it 
and compactifying it to four dimensions. We also show (\S~\ref{par5}) 
that the photon--axion mixing is described by the same formalism and 
lead to the same effects as for gravitons. Then, we turn to investigate 
the mixing in itself.  For that purpose, we sum up the known results of 
the mixing of a photon with a low mass particle in four dimensions 
(\S~\ref{par3_1}) and then discuss the most general case of a $D$ 
dimensional spacetime (\S~\ref{par3_2}).  The general expression for the 
oscillation probability is then evaluated in the particular cases of a 
five (\S~\ref{par3_3}) and of a six (\S~\ref{DD}) dimensional 
spacetime. We show that, as long as one is in a weak coupling regime, 
one can add the individual probabilities which leads to an enhancement 
of the oscillation probability if the KK modes are light enough. We also 
show that there exists a regime where the photon mixes strongly 
preferentially with a given KK mode.  We generalise our results to an 
inhomogeneous magnetic field (\S~\ref{par4}) when the probability of 
oscillation is small, and then turn to the cosmological and 
astrophysical situations where such effects may be observed.  We study 
the case of the cosmic microwave background (\S~\ref{par6_1}), of 
pulsars (\S~\ref{par6_2}) and of magnetars (\S~\ref{par5_3}). We show 
that even if the enhancement of the oscillation probability can be very 
important, it is still very difficult to observe this effect in known 
astrophysical systems. To finish (\S~\ref{par7}) we discuss laboratory 
experiments and particularly polarisation experiments. For that purpose, 
we describe the computation of the phase shift between the two 
polarisations of an electromagnetic wave and compare the result to the 
standard four dimensional case. 
 
\section{Equations of motion for the photon--graviton system}\label{par2}

Following \cite{giudice98,han98}, we consider a field theory defined by 
the $D$ dimensional action 
\begin{equation}\label{SD}  
  S_D=-\frac{1}{2\kappa^2_D}\int d^Dz\sqrt{-\bar g}\bar R +\int  
  d^Dz\sqrt{-\bar g}{\cal L}_m,  
\end{equation}  
where $\bar g_{AB}$ is the $D$ dimensional metric with signature 
$(-,+,...,+)$, $\kappa^2_D\equiv8\pi G_D=\bar M_D^{-(2+n)}$ and ${\cal 
L}_m$ is the matter Lagrangian. The indices $A,B,...$ take the value 
$0,..3,5..,D$ and we decompose $z^A$ as 
\begin{equation}\label{za}  
  z^A=(x^\mu,y^a)  
\end{equation}  
with $\mu,\nu,\ldots=0,\ldots,3$ and $a,b,\ldots=5,\ldots,D$.   
 
This theory is considered as being a low energy effective theory valid 
below some cut--off $M_{\rm max}$ in energy (see e.g. 
\cite{arkani98,arkani99,antoniadis98}). We will discuss in this paper 
only the cases $n=1$ and $n=2$ in details, and our conclusions, in these 
cases, are mostly cut--off independent.  We stress that the relationship 
between this cut--off and the fundamental string scale (if one wishes to 
embed these theories in superstring models) can be much more complicated than 
what is naively expected (see \cite{mourad99}). 
 
We expand the metric around the $D$ dimensional Minkowski spacetime as 
\begin{equation}\label{metric}  
  g_{AB}=\eta_{AB}+\frac{h_{AB}}{\bar M_D^{1+n/2}} 
\end{equation}  
where $\eta_{AB}$ is the $D$ dimensional Minkowski metric.  Inserting 
(\ref{metric}) in (\ref{SD}) and using the definition of the 
stress--energy tensor as $\sqrt{-\bar g}T_{AB}\equiv2\delta({\cal 
L}_m\sqrt{-\bar g})/\delta g^{AB}$, so that ${\cal L}_m\sqrt{-\bar 
g}={\cal L}_{m0}-h^{AB}T_{AB}/2\bar M_D^{1+n/2}$, we obtain the 
linearised action 
\begin{eqnarray}\label{lin_SD}  
  S_D&=&\int d^Dz\left[\frac{1}{2}h^{AB}\partial^C\partial_C  
  h_{AB}-\frac{1}{2}h^A_A\partial^C\partial_Ch_B^B\right.  
  +\frac{1}{2}h^{AB}\partial_A\partial_Bh^C_C  
  +\frac{1}{2}h^A_A\partial_C\partial_Bh^{CB}\nonumber\\  
  &&\qquad\qquad\left.-h^{AB}\partial_A\partial_Ch^C_B 
  -\frac{1}{\bar M_D^{1+n/2}}h^{AB}  
  T_{AB}+{\cal L}_{m0}\right].  
\end{eqnarray}  
We compactify this theory to get a four dimensional theory and use the 
periodicity on $y^a$ to expand the field $h^{AB}$ as 
\begin{equation}\label{decomposition}  
  h_{AB}(z^A)=\sum_{\vec{p}\in Z}\frac{h_{AB}^{(\vec  
  p)}(x^\mu)}{\sqrt{V_n}}\exp{(i\frac{p^ay_a}{R})}  
\end{equation}  
where $V_n=(2\pi R)^n$ is the volume of the compact $n$ dimensional 
space (assumed to be a cubic $n$--torus). $h_{AB}$ is split into a sum 
of KK modes living in the four dimensional spacetime. The ordinary 
matter being confined to the brane, its stress--energy tensor must 
satisfy 
\begin{equation}\label{tmunu}  
T_{AB}(z^C)=T_{\mu\nu}(x^\lambda)\delta^{(n)}(y^c)\eta^\mu_A\eta^\nu_B.  
\end{equation}  
In which follows, we restrict our attention to the case of an 
electromagnetic field $F_{\mu\nu}$ for which the stress--energy tensor 
is given by 
\begin{equation}\label{tdefmunu}  
  T_{\mu\nu}=F_{\mu\lambda}F_\nu^{\,\,\lambda}-\frac{1}{4}\eta_{\mu\nu}  
  F^{\lambda\rho}F_{\lambda\rho}.  
\end{equation} 
 
The fields $h_{AB}^{(\vec p)}$ can be decomposed into spin--2, spin--1 
and spin--0 four dimensional fields \cite{giudice98,han98}. Only spin--2 
and spin--0 particles couple to ordinary matter and spin--0 particles 
couple only to $T^\lambda_\lambda$. For the electromagnetic field 
$T^\lambda_\lambda=0$ classically so that the only relevant KK modes (at 
the tree level analysis of this article) will be the spin--2 
particles $G_{\mu\nu}^{(\vec p)}$ for which the action (\ref{lin_SD}) 
reduces to \cite{giudice98,han98} 
\begin{eqnarray}\label{lin_S4}  
  S_4&=&\int d^4x\left[\frac{1}{2}G_{\mu\nu}^{(\vec p)}(\Box-m^2_{\vec  
  p})G^{(\vec p)\mu\nu}+ G^{(\vec p)\mu\nu}\partial_\mu\partial_\lambda   
  G^{(\vec p)\lambda}_\nu\right.  
  -\frac{1}{2}G_{\mu}^{(\vec p)\mu}(\Box-m^2_{\vec  
  p})G^{(\vec p)\nu}_\nu  
  -G^{(\vec p)\mu\nu}\partial_\mu\partial_\nu G^{(\vec  
  p)\lambda}_\lambda \nonumber\\  
  &&\qquad\qquad-\left.\frac{1}{\bar M_4}G^{(\vec p)\mu\nu} 
  T_{\mu\nu}-\frac{1}{4}F^{\mu\nu}F_{\mu\nu}\right],  
\end{eqnarray}  
where $m^2_{\vec p}=\vec p^2/R^2$ is the square mass of the KK 
graviton, $\Box\equiv\partial_\mu\partial^\mu$. 
 
The equations of motion deduced from the Lagrangian (\ref{lin_S4}) are 
the coupled Einstein--Maxwell equations 
\begin{eqnarray}  
  &&(\Box-m^2_{\vec p})G^{(\vec p)}_{\mu\nu}=\frac{2}{\bar M_4}  
  T_{\mu\nu}\label{einstein1}\\  
  &&\partial^\mu G^{(\vec p)}_{\mu\nu}=0\label{einstein2}\\  
  &&G^{(\vec p)\mu}_{\mu}=0\label{einstein3}\\  
  &&\partial_\alpha F^{\alpha\beta}-\frac{2}{\bar  
  M_4}\sum_{\vec p}\partial_\alpha\left(G^{(\vec p)\alpha\nu}  
  F^\beta_\nu-G^{(\vec p)\beta\nu}  
  F^\alpha_\nu\right)=0.\label{maxwell1}  
\end{eqnarray}  
When $\vec p\not=0$, the field $G^{(\vec p)}_{\mu\nu}$ has 10-1-4=5 
degrees of freedom which is what is expected for a massive spin--2 
particle. 
  
We now consider an electromagnetic plane wave in the presence of a 
magnetic field $\vec H_0$ which is assumed constant on a  
characteristic scale $\Lambda_c$ in the sense that its variation  
in space and time are negligible on scales comparable 
to the photon wavelength and period. We define the basis 
\begin{equation}\label{base3d}  
  \vec e_\parallel\equiv\frac{\vec k}{k}, \quad  
  \vec e_\times\equiv\frac{\vec H_{0\perp}}{H_{0\perp}},\quad  
  \vec e_+,  
\end{equation}  
such that $(\vec e_\times,\vec e_+,\vec e_\parallel)$ is a direct 
orthonormal basis of the three dimensional space and where $\vec 
H_{0\perp}$ is the perpendicular component of $\vec H_0$ with respect to 
the direction of propagation $\vec k$.  We decompose the KK gravitons in 
scalar (S), vector (V) and tensor (T) as 
\begin{eqnarray}\label{SVT}  
  (S)&\quad& G_{00}^{(\vec p)}=\phi^{(\vec p)}, \quad G_{0i}^{(\vec p)}  
     =-ik_i{k^2}  
     \dot\phi^{(\vec p)}, \quad 
       G_{ij}^{(\vec p)}=\frac{\phi^{(\vec p)}}  
     {3}\delta_{ij}-\frac{3}{2k^2}  
      \Delta_{ij}\left(\frac{\phi}{3}  
     +\frac{\ddot\phi}{k^2}  
      \right)^{(\vec p)}\\  
  (V)&\quad& G_{0i}^{(\vec p)}\equiv V_i^{(\vec p)}=V_+^{(\vec p)}e_{i}^+  
     +V_\times^{(\vec p)} e_{i}^\times,\quad  
     G_{00}^{(\vec p)}=0,\quad G_{ij}^{(\vec  
     p)}=\frac{2}{k^2}k_{(j}\dot{V}_{i)}^{(\vec p)}\\  
  (T)&\quad& G_{00}^{(\vec p)}=0,\quad G_{0i}^{(\vec p)}=0,  
      \quad G_{ij}^{(\vec p)}=G_+^{(\vec p)}\epsilon_{ij}^+ 
     +G_\times^{(\vec p)}\epsilon_{ij}^\times,  
\end{eqnarray}  
where $\Delta_{ij}\equiv\left(k_ik_j-\frac{k^2}{3}\delta_{ij}\right)$, 
$i,j=1..3$ and a dot refers to a time derivative. The polarisation 
tensor of the graviton modes 
$\epsilon_{ij}^\lambda$ is defined by 
\begin{eqnarray}\label{polarisation}  
  \epsilon_{ij}^\lambda\equiv\left(e_i^\times e_j^\times - e_i^+  
  e_j^+\right)\delta^\lambda_\times +  
  2e_i^{(+}e_j^{\times)}\delta^\lambda_+.  
\end{eqnarray}  
The advantage of such a decomposition is that the scalar, vector and 
tensor contributions decouple. The five degrees of freedom of each 
massive spin--2 KK gravitons have been decomposed in one scalar mode 
($\phi^{(\vec p)}$), two vector modes ($V_{+/\times}^{(\vec p)}$) and 
two tensor modes ($G_{+/\times}^{(\vec p)}$). Each of these modes 
satisfies independently the constraints 
(\ref{einstein2}--\ref{einstein3}). 
  
We consider an electromagnetic wave with a potential vector of the 
form 
\begin{equation}\label{emwave}  
  \vec A=i(A_\times(u),A_+(u),0)\hbox{e}^{-i\omega t},  
\end{equation}  
where $u$ is the coordinate along the direction of propagation. We 
have introduced the arbitrary phase $i$ so that its electric and 
magnetic fields are 
\begin{eqnarray}  
  \vec E&\equiv&-\partial_t\vec A  
  =(\omega A_\times(u),\omega A_+(u),0)\hbox{e}^{-i\omega t}  
  \label{defE}\\  
  \vec B&\equiv& \hbox{curl}(\vec A)=(-i\partial_u  
  A_+(u),i\partial_uA_\times(u),0)\hbox{e}^{-i\omega t}\label{Bwave}.  
\end{eqnarray}  
The stress--energy tensor of these waves in the presence of $\vec H_0$ 
has no vector component. Its tensor component is given by 
\begin{equation}\label{tt}  
  T_{ij}=i\sum_{\lambda=+,\times}\partial_uA_\lambda 
  H_{0\perp}\hbox{e}^{-i\omega t}\epsilon_{ij}^\lambda. 
\end{equation}  
We see, as expected, that a plane wave possesses a tensor part only if 
it propagates in an external field and that the polarisations $+$ and 
$\times$ of the electromagnetic wave couple respectively to the 
polarisations $+$ and $\times$ of the gravitons. The electromagnetic 
wave generates also scalar perturbations, but it can be shown 
\cite{magueijo94} that (in the usual four dimensional case) the total 
energy converted in this scalar wave are negligible compared to the 
tensor contribution. In the following, we concentrate on the tensor 
modes. 
  
The equation of evolution of this system is given by the Einstein 
equation (\ref{einstein1}) which reduces to 
\begin{equation}\label{einsteinfinal}  
  \left(\omega^2+\partial_u^2-m^2_{\vec p}\right)G^{(\vec  
  p)}_\lambda=\frac{2iH_{0\perp}}{\bar M_4} \partial_u A_\lambda,  
\end{equation}  
and the Maxwell equation (\ref{maxwell1}) which reduces to 
\begin{equation}\label{maxwellfinal}  
  \left(\omega^2+\partial_u^2\right)A_\lambda=\frac{2iH_{0\perp}}{\bar  
  M_4}\sum_{\vec p}  
  \partial_u G_\lambda^{(\vec p)},  
\end{equation}  
where we have used the ansatz $G_{ij}=\sum_\lambda 
G_\lambda(u)\hbox{e}^{-i\omega t}\epsilon_{ij}^\lambda$ for the 
gravitons. 
  
Since we have assumed that the magnetic field varies in space on 
scales much larger than the photon wavelength, we can perform the 
expansion 
$\omega^2+\partial_u^2=(\omega+i\partial_u)(\omega-i\partial_u) 
=(\omega+k)(\omega-i\partial_u)$ for a field propagating in the $+u$ 
direction. If we assume a general dispersion equation of the form 
$\omega=nk$ and that the refractive index $n$ satisfies $|n-1|\ll1$, 
we may approximate $\omega+k=2\omega$ and $k/\omega=1$.  This 
approximation can be understood as a WKB limit where we set 
$A(u)=|A(u)|\hbox{e}^{iku}$ and assume that the amplitude $|A|$ varies 
slowly, i.e. that $\partial_u |A|\ll k|A|$. In that limit, the system 
(\ref{einsteinfinal}--\ref{maxwellfinal}) reduces to 
\begin{equation}\label{systfinal}  
  \left[\omega-i\partial_u+{\cal M}_\lambda\right]  
  \left[\begin{array}{c}  
        A_\lambda\\  
        G_\lambda^{(0)}\\  
        \vdots\\  
        G_\lambda^{(q)}\\  
        \vdots  
        \end{array}  
  \right]=0,  
\end{equation}  
the matrix ${\cal M}_\lambda$ being given by  
\begin{equation}\label{M}  
  \left(\begin{array}{ccccccc}  
  \Delta_\lambda & \Delta_M &\Delta_M&\cdots&\Delta_M&\cdots&\cdots\\  
  \Delta_M& \Delta_m^{(0)}&0&\cdots  & 0&\cdots &\cdots\\  
  \Delta_M& 0 & \Delta_m^{(1)}&0&0&\cdots&\cdots \\  
  \vdots&\vdots&\ddots&\ddots&\ddots&\ddots&\cdots\\  
  \Delta_M&0&\cdots&0&\Delta_m^{(q)}& 0&\ddots \\  
  \vdots&\vdots&\vdots&\vdots&\ddots&\ddots&\ddots  
  \end{array}\right)  
\end{equation}  
with  
\begin{equation}\label{deltas}  
  \Delta_M\equiv\frac{H_{0\perp}}{\bar M_4}\quad\hbox{and}\quad  
  \Delta_m^{(q)}\equiv \vec p^2_{(q)} \Delta_m.  
\end{equation}  
$\vec p_{(q)}$ is a n--uplets $(p_{q_1},p_{q_2},..p_{q_n})$ of integers 
and we have ordered the $\Delta_m^{(q)}$ such that 
$$\left|\Delta_m^{(q)}\right|\leq\left|\Delta_m^{(q+1)}\right|$$  
and $\Delta_m$ is defined by  
\begin{equation}  
  \Delta_m\equiv\frac{-1}{2R^2\omega}.  
\end{equation}  
Each $\Delta_m^{(q)}$ appears with a multiplicity given by the  
number of n--uplets having the same norm $\sum_{i=1..n} p_{q_i}^2$.  
We define the two series $(r_i)_{i\geq1}$ and $(s_i)_{i\geq1}$   
such that  
\begin{equation}\label{rang}  
\Delta_m^{(r_i-1)}<\Delta_m^{(r_i)}=\Delta_m^{(r_i+1)}=\ldots  
\Delta_m^{(r_i+s_i-1)}  
<\Delta_m^{(r_i+s_i)}\equiv\Delta_m^{(r_{i+1})}.  
\end{equation}  
We have $r_{i+1} = r_i + s_i$, and $s_i$ is the multiplicity of the 
element $\Delta_m^{(r_i)}$, i.e.  the number of times it appears in 
the matrix (\ref{M}).  $r_i$ is the rank in the series 
$(\Delta_m^{(0)},\Delta_m^{(1)},\ldots)$ where the $i^{\rm th}$ 
distinct value of $\Delta_m^{(q)}$ appears for the first time. In the 
case of a five dimensional spacetime one can easily find out 
that $s_1=1$ and $s_i=2$ for $i>1$  and that $r_1=0, r_2=1, r_3=3, 
\ldots$ In the case of a six dimensional spacetime, $s_i = 
(1,4,4,\ldots)$, $r_i=(0,1,5,9,\ldots)$.  Introducing the cut--off 
$M_{\rm max}$ discussed above, we require $m_{\vec{p}}^2=\vec{p}^2/R^2 
<M^2_{\rm max}$, which using (\ref{planck}) translates into $\vec{p}^2 < 
p_{\rm max}^2$ with 
\begin{equation}  
 p_{\rm max} = \left(\frac{\bar M_4}{M_D}\right)^{2/n} 
\left(\frac{M_{\rm max}}{M_D}\right).     
\end{equation}  
Setting $M_{\rm max} \sim M_D$, we obtain 
$p_{\rm max}\sim\left(\bar{M}_4/M_{D} \right)^{2/n}$. 
For $n=2$ and $M_D \sim 1 {\rm TeV}$ one finds  
\begin{equation} 
p_{\rm max} \sim 10^{15} \label{pmax2d}, 
\end{equation} 
which means that we have to consider a very large number of KK states. 
We also define a maximum index, $N$ say, for the series $\Delta_m^{(q)}$ 
defined by 
\begin{equation}  
N\equiv\mbox{sup}\{ q\quad|\quad \vec p^2_{(q)}=p^2_{\rm max}\},  
\end{equation}  
which translates into a maximum index $N_D$ for the series  
$s_{i}$ and $r_i$. We stress that the number of KK modes relevant for 
the photon-- KK graviton oscillation is likely to be smaller than $N$ 
due to decoherence effects, such as the source and detector finite 
width in momentum, the wave packet separation for massive (and  
non--relativistic) KK modes... (see e.g. \cite{decohe} for a description 
of these effects in the case of neutrino oscillation).

The term $\Delta_\lambda$ can be decomposed as 
$\Delta_\lambda=\Delta_{\rm QED}+\Delta_{\rm CM}+\Delta_{\rm plasma}$. 
The first term contains the effect of vacuum polarisation giving a 
refractive index to the photon (see e.g. Adler \cite{adler71}) and can 
be computed by adding the Euler--Heisenberg effective Lagrangian which 
is the lowest order term of the non--linearity of the Maxwell equations 
in vacuum (see e.g. \cite{euler36,iz}) to the action (\ref{lin_S4}) 
\footnote{The equation of motion derived from (\ref{lin_S4}) is 
(\ref{systfinal}) with $\Delta_\lambda=0$. We intentionaly omit the 
Euler--Heisenberg contribution in the presentation for the sake of 
clarity. Its Lagrangian is explicitely given by ${\cal L}_{EH}= 
\frac{\alpha^2}{90m_e^4}\left[(F^{\mu\nu}F_{\mu\nu})^2+\frac{7}{4} 
(F^{\mu\nu}\tilde F_{\mu\nu})^2\right]$.}.  The second term describes 
the Cotton--Mouton effect, i.e. the birefringence of gases and liquids 
in presence of a magnetic field and the third term the effect of the 
plasma (since, in general, the photon does not propagate in 
vacuum). Their explicit expressions are given by 
\begin{eqnarray}\label{deltabis}  
  &&\Delta_{\rm QED}^\times=\frac{7}{2}\omega\xi,\quad  
  \Delta_{\rm QED}^+=2\omega\xi,\nonumber\\  
  &&\Delta_{\rm plasma}=-\frac{\omega_{\rm plasma}^2}{2\omega},\nonumber\\  
  &&\Delta_{\rm CM}^\times-\Delta_{\rm CM}^+=2\pi CH_0^2,  
\end{eqnarray}  
with $\xi\equiv(\alpha/45\pi)(H_{0\perp}/H_c)^2$, $H_c\equiv 
m_e^2/e=4.41\times10^{13}\,\hbox{G}$, $m_e$ the electron mass, $e$ the 
electron charge and $\alpha$ the fine structure constant.  $C$ is the 
Cotton--Mouton constant \cite{cotton}; this effect gives only the 
difference of the refractive indices and the exact value of $C$ is 
hard to determine \cite{cotton2}; we will neglect this effect but for 
the polarisation experiments (see \S~\ref{par7_2}).  The plasma 
frequency $\omega_{\rm plasma}$ is defined by 
\begin{equation}  
\omega_{\rm plasma}^2\equiv 4\pi\alpha\frac{n_e}{m_e},  
\end{equation}  
$n_e$ being the electron density. Note that the $\Delta_m^{(q)}$ are  
always negative whereas $\Delta_\lambda$ is positive if the  
contribution of the vacuum dominates and negative when the plasma  
term dominates.

The equation of motion (\ref{systfinal}) reduces to the one studied by 
Raffelt and Stodolsky \cite{raffelt88} when one considers only four 
dimensions so that ${\cal M}_\lambda$ contains only the massless 
graviton and is then a $2\times2$ matrix. The main difference lies in 
the fact that now the electromagnetic component couples to a large 
numbers of KK gravitons.  This can be compared to some models of 
neutrino oscillations in spacetime with extra--dimensions 
\cite{arkani98b,dudas}. We should also note that the two polarisations 
are, as expected, completely decoupled and obey the same equation of 
evolution. In the following of this article, we omit the subscript 
$\lambda$ of the polarisation. 
 
\section{Equations of motion for the photon--axion system}\label{par5}  
  
Before turning to the study the photon--graviton mixing, we consider the 
case of axions and show that the photon--axion mixing can be described 
by the same formalism. 
 
We consider the generic action \cite{chang99,dudas2} for the bulk axion 
 photon system 
\begin{eqnarray}\label{S4axion}  
  S_4&=&\int d^4x\left[\sum_{\vec p}\left(-\frac{1}{2}\left\lbrace  
  \partial^\mu a^{(\vec p)}\partial_\mu a^{(\vec p)}+m^2_{\vec p}  
  {a^{(\vec p)}}^2\right\rbrace\right.\right. 
  \left.\left.+\frac{a^{(\vec p)}}{f_{\rm PQ}}F_{\mu\nu}  
     \widetilde F^{\mu\nu}\right)-\frac{1}{4}F_{\mu\nu}F^{\mu\nu}\right],  
\end{eqnarray}  
where the $a^{(\vec p)}$ are the mass eigenstates of the axions and 
$m_{\vec p}$ their masses. $\widetilde F_{\mu\nu}\equiv 
\frac{1}{2}\epsilon_{\mu\nu\rho\sigma}F^{\rho\sigma}$ is the dual of the 
electromagnetic tensor, $\epsilon_{\mu\nu\rho\sigma}$ being the 
completely antisymetric tensor such that $\epsilon_{0123}=+1$. As for 
the gravitons, the mass spectrum is expected to be discrete, the states 
can be labelled by a n-uplet $\vec p$ and is expected to have a typical 
spacing of $1/R$. We have considered here that every axion KK state 
$a^{(\vec p)}$ couples to the photon with the same coupling $1/f_{\rm 
PQ}$. This is only expected to be true if the typical mass, $m_{\rm 
PQ}$, given to the axion zero mode by instanton effects is much lower 
than the typical KK mass $1/R$ \cite{dudas2}. Let us further stress here 
that for such bulk axions the usual relationship between the axion mass 
and the PQ scale does not hold anymore, so that one expects to see 
interesting new effects to appear \cite{dudas2}. Inspired by the usual 
bounds on $f_{\rm PQ}$, we take $f_{\rm PQ}$ of order 
$10^{10}$~GeV. However we stress that the usual bounds on $f_{\rm PQ}$ 
may be modified partly because of a large number of axion--like 
particles coupling to the standard model fields. For example, we expect 
that the astrophysical bounds will be more stringent mainly because a 
star will now be able to emit all the energetically accessible modes 
(see \cite{chang99} and also \cite{dudas2} for a discussion on relic axions 
oscillations).  
 
We do not consider the perturbations of the metric and work in Minkowski 
spacetime since we are interested in the interaction between the photon 
and the axion.  We deduce from (\ref{S4axion}) the coupled Klein--Gordon 
and Maxwell equations 
\begin{eqnarray}  
&&\left(\Box-m^2_{\vec p}\right)a^{(\vec p)}=-\frac{1}{f_{\rm PQ}}  
  F_{\mu\nu}\widetilde F^{\mu\nu},\\  
&&\partial_\alpha F^{\alpha\beta}=\frac{4}{f_{\rm PQ}}\partial_\alpha  
  \left[\sum_{\vec p} a^{(\vec p)}\widetilde F^{\alpha\beta}\right].   
\end{eqnarray}  
We now decompose the electromagnetic wave as in (\ref{emwave}-\ref{Bwave})   
with respect to the basis (\ref{base3d}), so that the former  
system reads 
\begin{eqnarray}  
&&\left(\Box-m^2_{\vec p}\right)a^{(\vec p)}=\frac{4H_{0\perp}}{f_{\rm PQ}}  
A_\times\\  
&&\Box A_\lambda=\frac{4H_{0\perp}}{f_{\rm PQ}}\omega\delta_{\lambda\times}  
\sum_{\vec p}a^{(\vec p)},   
\end{eqnarray}  
where we have decomposed the axions as $a^{(\vec p)}(u)\exp{(-i\omega 
t)}$.  Using the same WKB limit as in section \ref{par3}, we obtain the 
linearised system 
\begin{equation}  
\left(\omega-i\partial_u+{\cal M}\right)  
\left[  
\begin{array}{ccc}  
A_+\\A_\times\\a^{(\vec 0)}\\ \vdots\\a^{(\vec p)}\\ \vdots  
\end{array}  
\right]=0.  
\end{equation}  
The matrix ${\cal M}$ is now defined by  
\begin{equation}  
{\cal M}=\left(  
\begin{array}{cccccc}  
\Delta_+&0&0&0&0&\cdots\\  
0&\Delta_\times&\Delta_M&\Delta_M &\cdots\\  
0&\Delta_M&\Delta_a^{(0)}&0&0&\cdots\\  
0&\Delta_M&0&\ddots&\ddots&\ddots\\  
\vdots&\vdots&\vdots&\ddots&\Delta_a^{(q)}&\ddots\\  
\vdots&\vdots&\vdots&\vdots&\ddots&\ddots  
\end{array}  
\right)  
\end{equation}  
with  
\begin{equation}  
\Delta_M=\frac{4H_{0\perp}}{f_{\rm PQ}},  
\quad  
\Delta_a^{(\vec p)}=-\frac{m^2_{\vec p}}{2\omega},  
\end{equation}  
$\Delta_+$ and $\Delta_\times$ being given by equation (\ref{deltabis}). 
This system reduces to the Raffelt and Stodolsky \cite{raffelt88} system 
when we consider only four dimensions. Only the component $\times$, 
i.e. parallel to the magnetic field, couples to the axions. This is a major 
difference compared with gravitons for which both polarisations of the photon 
evolve alike whereas here only $A_\times$ is affected by the mixing. 
  
One of the goal of this section was to set the theoretical framework for 
further experimental studies of photon--bulk axion oscillations (see 
\S~\ref{par7} for a more detailed discussion) and to show that it is 
described by a similar formalism as photon--KK graviton oscillations (under the
validity conditions explained below equation (\ref{S4axion}). 
 
\section{Photon--KK state mixing in a homogeneous field}\label{par3}  
  
We now describe the physical implications of the system 
(\ref{einsteinfinal}--\ref{maxwellfinal}) and start by reviewing briefly 
the well studied problem of the mixing of a photon with a low mass 
particle in four dimensions (\S~\ref{par3_1}).  We then give the exact 
expression of the oscillation probability in $D$ dimensions and discuss 
qualitatively its magnitude and the effect of the coupling of the photon 
to a large number of particles. 
 
\subsection{The usual photon mixing  with a low mass particle}\label{par3_1}  
  
This case was well studied in the literature, see e.g.  Raffelt and 
Stodolsky \cite{raffelt88} and we just summarize the main features of 
the results to compare to the case of a spacetime with 
extra--dimensions.  For the mixing with a single particle of mass $m$, 
the matrix ${\cal M}$ reduces to 
\begin{equation}  
  {\cal M}=\left(  
  \begin{array}{cc}  
    \Delta_\lambda&\Delta_M\\  
    \Delta_M&\Delta_m  
  \end{array}  
  \right) 
\end{equation} 
with $\Delta_m\equiv-m^2/2\omega$. 
The solution to the equation of motion (\ref{systfinal}) 
is obtained by diagonalising ${\cal M}$ throught a rotation 
\begin{equation}\label{rotation}  
  \left[\begin{array}{c}A'\\G'\end{array}\right]=  
  \left(\begin{array}{cc}\cos\vartheta&\sin\vartheta\\  
  -\sin\vartheta&\cos\vartheta\end{array}\right)  
  \left[\begin{array}{c}A\\G\end{array}\right]  
\end{equation}  
with 
\begin{equation}\label{theta}  
  \tan2\vartheta\equiv2\frac{\Delta_M}{\Delta_\lambda-\Delta_m}  
   =\frac{2\alpha}{1-\beta} 
\end{equation}  
and where we have introduced $\alpha\equiv\Delta_M/\Delta_\lambda$ and   
$\beta\equiv\Delta_m/\Delta_\lambda$. We obtain by solving 
(\ref{systfinal}) in this new basis  
\begin{eqnarray}  
  A'(u)&=&\hbox{e}^{-i\Delta_\lambda'u}A'(0)\nonumber\\  
  G'(u)&=&\hbox{e}^{-i\Delta_g'u}G'(0)  
\end{eqnarray}  
where a global phase $\omega u$ has been omitted. 
The two eigenvalues $\Delta_\lambda'$ and $\Delta_g'$  
of ${\cal M}$ are explicitly given by  
\begin{eqnarray}  
  \Delta_\lambda'=\frac{\Delta_\lambda+\Delta_m}{2}+  
  \frac{\Delta_\lambda-\Delta_m}{2 \cos2\vartheta}\quad{\rm and}\quad  
   \Delta_g'=\frac{\Delta_\lambda+\Delta_m}{2}-  
  \frac{\Delta_\lambda-\Delta_m}{2 \cos2\vartheta}.\label{defDelta}  
\end{eqnarray}  
Going back to the initial basis, we obtain  
\begin{eqnarray} \label{53} 
  A(u)&=&\left(\hbox{e}^{-i\Delta_\lambda'u}\cos^2\vartheta+  
         \hbox{e}^{-i\Delta_g'u}\sin^2\vartheta\right)A(0)  
       +\sin\vartheta\cos\vartheta\left(\hbox{e}^{-i\Delta_\lambda'u}-  
         \hbox{e}^{-i\Delta_g'u}\right)G(0),\nonumber\\  
G(u)&=&\sin\vartheta\cos\vartheta\left(\hbox{e}^{-i\Delta_\lambda'u}-  
         \hbox{e}^{-i\Delta_g'u}\right)A(0)  
      +\left(\hbox{e}^{-i\Delta_g'u}\cos^2\vartheta+  
         \hbox{e}^{-i\Delta_\lambda'u}\sin^2\vartheta\right)G(0).  
\end{eqnarray}  
The oscillation probability of a photon into a graviton is  
computed by considering the initial state $(A(0)=1,G(0)=0)$ and is given  
by  
\begin{eqnarray}  
  P(\gamma\rightarrow g)\equiv\mid\langle A(0)\mid G(u)  
  \rangle\mid^2 
  &=&\sin^2\left(2 \vartheta \right)\sin^2\left( 
  \frac{\Delta_{\rm osc}}{2}u\right),\label{p57}\\  
  &=&\left(\Delta_M u\right)^2 {\sin^2(\Delta_{\rm osc} u /2) 
  \over (\Delta_{\rm osc} u /2)^2}  
\end{eqnarray}  
with  
\begin{equation}  
\Delta_{\rm osc}\equiv  
\Delta_\lambda'-\Delta_g'=\frac{\Delta_\lambda-\Delta_m}{\cos 2\vartheta}  
=\frac{2\Delta_M}{\sin2\vartheta}=\frac{1-\beta}{\cos2\vartheta} 
\Delta_\lambda 
\end{equation}  
The oscillation length is thus given by $\ell_{\rm 
osc}\equiv2\pi/\Delta_{\rm osc}$.  We see that the oscillation 
probability cannot be larger than $\left( \Delta_M u \right)^2$. One has 
to be aware that $\vartheta$ depends on the polarisation index $\lambda$. 
  
It is interesting to single out the two following limiting regimes:  
\begin{itemize}  
  \item The {\it weak mixing} regime in which $\vartheta\ll1$ so  
  that the probability (\ref{p57}) reduces to  
\begin{equation}\label{weakP}  
  P(\gamma\rightarrow g)=4{\alpha^2 \over \left( 1-\beta\right)^2}  
  \sin^2\left(\frac{1-\beta}{2} \Delta_\lambda u\right).  
\end{equation}  
When the oscillation length $\ell_{\rm 
osc}=\frac{2\pi\vartheta}{\Delta_M}$ is large with respect to the 
coherent path distance $u$, the weak mixing probability can be further 
approximated (with $\Delta_M u \ll \vartheta \ll 1$) by 
\begin{equation}  
  P(\gamma\rightarrow g)\simeq\left(\alpha \Delta_\lambda u \right)^2\simeq 
  \left(\Delta_M u\right)^2.  
\end{equation}  

\item The {\it strong mixing} regime in which the mixing is  
  maximal, i.e. when $\vartheta\simeq\pi/4$, so that the oscillation 
probability reduces to  
\begin{equation}\label{strongP}  
  P(\gamma\rightarrow g)=\sin^2\left(\Delta_Mu\right)  
\end{equation}  
and the oscillation length to  
\begin{equation}  
\ell_{\rm osc}=\frac{\pi}{\Delta_M}. \label{e57} 
\end{equation}  
A complete transition between a photon and the light particle is then 
possible. This can only happen when $\Delta_m$ and $\Delta_\lambda$ have 
the same sign (see equation (\ref{theta})).  We further note here that 
the width in $\beta$ of the strong mixing region is roughly given by 
$\alpha$ according to equation (\ref{theta}). 
\end{itemize}

\subsection{Mixing in $D$ dimensions}\label{par3_2}  
 
\subsubsection{General result} 
  
To compute the oscillation probability in a spacetime with 
extra--dimensions, we first have to solve (\ref{systfinal}) which 
implies the diagonalisation of the matrix (\ref{M}). We present the 
explicit and detailed computation of both the eigenvalues and 
eigenvectors in appendix \ref{appA}.  We then compute in appendix 
\ref{appB} the explicit form of the oscillation probability (see 
equation (\ref{B5})) 
\begin{equation}\label{pp}  
P(\gamma\rightarrow g)=1-\left|\sum_{i=1}^{N_D}f^2_{x_i}  
\hbox{e}^{ix_iu}\right|^2.  
\end{equation}  
Taking into account the fact that $\sum f_{x_i}^2=1$,  
it can be rewritten as  
\begin{equation}\label{ppp}  
P(\gamma\rightarrow g)=2\sum_{i,j=1}^{N_D}f_{x_i}^2f_{x_j}^2  
\sin^2\left[\frac{x_i-x_j}{2}u\right],  
\end{equation}  
where the coefficients $f^2_{x_j}$ are defined by (see equation (\ref{B6})) 
\begin{equation}\label{f2}  
f_{x_j}^2\equiv\left[1+ \alpha^2\sum_{i=1}^{N_D}\frac{s_i} 
{\left(y_j-\beta_i\right)^2} \right]^{-1}.  
\end{equation} 
The expressions (\ref{ppp}) and (\ref{f2}) depend on the eigenvalues 
$y_i$ solutions of the equation (\ref{eqxi}). Introducing the 
notations $y\equiv x/\Delta_\lambda$, 
$\alpha\equiv\Delta_M/\Delta_\lambda$, 
$\beta\equiv\Delta_m/\Delta_\lambda$ and 
$\beta_i\equiv\Delta^{(r_i)}_m/\Delta_\lambda$, the eigenvalues 
equation (\ref{eqxi}) can be rewritten as 
\begin{equation}\label{y}  
y-1= \alpha^2\sum_{i=1}^{N_D}\frac{s_i}{y-\beta_i}. 
\end{equation}  
The photon--KK graviton oscillations is then completely described 
by the set of equations (\ref{ppp}--\ref{y}). 
 
Indeed, it is difficult (even impossible if $n >1$) to compute 
analytically the roots of (\ref{y}). For instance the coefficients 
$s_i$ are not known analytically if $n>1$; it is of course possible to 
compute $P(\gamma\rightarrow g)$ numerically, but this is not our 
purpose here.  In the next two sections we derive the oscillation 
probability in the two cases $n=1$ and $n=2$ in a range of parameters 
dictated by the systems where such a mixing may appear.  In the next 
paragraph, we discuss qualitatively the results found there, stressing 
some new effects due to the presence of a large number of KK states, as 
well as to the degeneracy of each KK level. 
 
\subsubsection{Qualitative discussion } \label{qual} 
 
We only discuss the cases where $\alpha$ is small in comparison to 
$\beta$ as dictated by the physical systems studied in \S~\ref{par6} and 
\S~\ref{par7}.  Two different limiting regimes appear, a {\it large 
radius} regime (when $|\beta|$ is smaller than unity) and for which 
there is a significant effect of the extra--dimensions, and a {\it small 
radius} regime (when $|\beta|$ is larger than unity) and for which there 
is only small departure from the usual four dimensional photon mixing.\\ 
 
Let us first discuss the large radius regime where $|\beta|$ is smaller 
than unity.  As in four dimensions, according to the respective value of 
$\Delta_\lambda$ and of the $\Delta_m^{(q)}$'s, two behaviours can 
appear: 
\begin{itemize} 
\item A {\it strong mixing} regime with one given KK state , if there exists  
a state $K$ such that  
\begin{equation}\label{sk} 
|\Delta_m^{(K)}-\Delta_\lambda|\ll\sqrt{s_K}\Delta_M. 
\end{equation} 
This can happen only if $\beta>0$, i.e. when plasma effects dominate 
over the vacuum polarisation. We stress also that since we have assumed 
along this discussion that $\alpha$ is lower than $\beta$ there is at 
most one KK state which can mix strongly with the photon. The total 
probability will be found to be dominated by a term of the form 
\begin{equation}\label{E63} 
P(\gamma\rightarrow g)=(1-\eta)\sin^2\left[\Delta_{\rm osc}^{(K)} 
u\right]+4\sum_{i\not=K} \frac{s_i \alpha^2}{(1-\beta_i)^2} 
\sin^2\left(\frac{1-\beta_i}{2} \Delta_\lambda u \right), 
\end{equation} 
with 
\begin{equation}\label{E64} 
\Delta_{\rm osc}^{(K)} =\sqrt{s_K} \Delta_M 
\end{equation} 
(see \S~\ref{par3_3} and \S~\ref{DD} for a detailed derivation). 
$\eta$ is much smaller than unity. As will be shown later, this 
form accounts for keeping only the dominant part of each 
$f^2_{x_i}$. Depending on the argument of the sines, the probability is 
either dominated by $\sin^2\left[\Delta_{\rm osc}^{(K)}u\right]$ 
or by the correction coming from the modes $i\not=K$. This 
shows a first departure to the four dimensional case due to the 
degeneracy of the KK level $K$; the oscillation length associated with 
the strong mixing state (labelled by $K$) is lowered by a factor 
$\sqrt{s_K}$ which can be very large. Moreover, the width of the region 
in $\Delta_\lambda$ of strong mixing is larger by a factor $\sqrt{s_K}$ 
than in the usual case [see below equation (\ref{e57})]. 

An other important difference with the usual four dimensional situation, 
where the strong mixing regime can only occur when $\Delta_\lambda$ 
crosses {\it the unique} $\Delta_m$ characteristic of the mixing state, 
we now have more possibilities to be in that regime, where a complete 
transition between the photon and the graviton is possible. Because of 
the presence of a KK state $\Delta_m^{(q)}$ in any interval in 
$\Delta_\lambda$ of typical width $\Delta_m$, only {\it fluctuations} of 
$\Delta_\lambda$ of order $\Delta_m$ can lead to it.

\item A {\it weak mixing} regime where for all $q$, 
$|\beta_q-1|\gg\alpha$.  The oscillation probability is then given by 
\begin{equation} \label{E65} 
P(\gamma\rightarrow g)\simeq 4\sum_{i} \frac{s_i \alpha^2}{(1-\beta_i)^2} 
 \sin^2\left(\frac{1-\beta_i}{2}  
 \Delta_\lambda u \right). 
\end{equation} 
This contribution is exactly the one that will be intuitively thought of 
and obtained by summing the individual oscillation probabilities 
(\ref{weakP}) of the photon into each KK state with the mixing angle 
\begin{equation}  
\tan2\vartheta_q\equiv\frac{2\alpha}{1-\beta_q}. \label{thetai}  
\end{equation}

There are roughly three contributions to the sum (\ref{E65}) that we 
estimate as follows. 
 
\begin{enumerate} 
\item All the states such that $|\beta_q|\ll1$ mix with the photon with 
approximatively the same angle $\vartheta_q\sim\alpha$ if we neglect 
$\beta_q$ with respect to unity in (\ref{thetai}). The order of 
magnitude of the probability of oscillations with these states is then 
\begin{equation} 
P(\gamma\rightarrow g)\sim 4{\cal N}_1\alpha^2\sin^2 
\left(\frac{\Delta_\lambda}{2}u\right). 
\end{equation} 
${\cal N}_1$ can be estimated by counting the number of modes such that 
$\beta_q\leq\beta_{N_1}\simeq1$ with their multiplicity (see appendix 
\ref{appD}), i.e.  ${\cal N}_1\sim \sum_{k=1}^{N_1} 
k^{n-1}\sim N_1^n\sim\beta^{-n/2}$ so that 
\begin{equation} 
P(\gamma\rightarrow g)\sim \frac{\alpha^2}{\beta^{n/2}} \sin^2 
\left(\frac{\Delta_\lambda}{2}u\right). 
\end{equation} 
This already shows that the oscillation probability can be greatly 
enhanced (by a factor $\beta^{-n/2}$ with respect to the four 
dimensional case with the same mixing parameters [obtained from equation 
(\ref{weakP}) with $|\beta| \ll 1$]). 
 
\item All $\beta_q$ such that $|\beta_q|\gg1$ have a mixing angle 
roughtly estimated by $\vartheta_q\sim\alpha/\beta_q$ and their 
contribution to the probability is of order 
\begin{equation} 
P(\gamma\rightarrow g)\sim 4 \alpha^2 
\sum_{\beta_q>1}\frac{1}{\beta_q^2} 
\sin^2\left(\frac{\Delta_\lambda\beta_q}{2}u\right). 
\end{equation} 
This series is difficult to evaluate since the oscillation length is 
different for each KK state.  When $n\leq3$ it can be bounded by 
$\alpha^2/\beta^{n/2}$ so that this contribution is at most of the same 
order of magnitude than the previous one. 
 
\item The contribution of the $\beta_q$ such that $\beta_q\sim1$ which 
only exists if $\beta>0$ is bounded by 
$\alpha^2\sum_{\beta_i\sim1}s_i/(\beta_i-1)^2$. First of all, since 
$|\beta_q-1|\gg\alpha^2$ we are never in a strong mixing regime. Now, we 
single out $\beta_K$, the closest $\beta_i$ to unity from which it 
follows that $\forall i\not=K$, $|\beta_i-1|\geq\beta/2$ and thus  
the contribution of all the $\beta_i\sim1$ for $i\not=K$ is bounded by 
$(\alpha^2/\beta^2)\sum_{\beta_i\sim 1,i\not= K} s_i$. It  can be 
dominated by the contribution of the term $K$ given by $\alpha^2s_K/ 
|\beta_K-1|^2$ according to the relative value of $|\beta_K-1|$ in units 
of $\beta$. 
\end{enumerate} 
 
In conclusion, the weak mixing case is characterised by an enhancement 
of the probability by a factor at least $\beta^{-n/2}$  due to the 
fact that the photon couples to a large number of KK states. We further
note that when $\beta<0$ and $|\beta|\ll1$ one can obtain an absolute
bound on the oscillation probability of order $\alpha^2/\beta^2$
for $n\leq3$ (when $n=2$, this bound is given by
$10{\cal Q}\alpha^2/\beta^2$ [see appendix~(\ref{appD})]).
\end{itemize} 
 
We now turn to the {\it small radius} regime where $|\beta|\gg1$ and in  
which the photon mixes preferentially with the zero mode. The 
probability (\ref{ppp}) can be expressed as 
\begin{equation} 
P(\gamma\rightarrow g)=(1-\epsilon)P_{4D}(\gamma\rightarrow g) 
+4\sum_{i>1}\frac{\alpha^2}{(1-\beta_i)^2} 
\sin^2\left(\frac{\Delta_\lambda(1-\beta_i)}{2}u\right), 
\end{equation} 
where $P_{4D}$ is the oscillation probability for the mixing with the 
zero mode and is given by (\ref{p57}) and $\epsilon\ll1$ is the 
correction of the oscillation probability with this mode coming from the 
existence of the extra--dimensions. In this case, the lightest massive 
KK mode is so heavy compared to the photon effective mass that it can 
barely be excited by the photon. The contribution of the other KK 
modes can be shown to be bounded by ${\cal O}(\alpha^2\beta^{-2})$. 
The contribution of the massive KK states is {\it suppressed} by a 
factor $\beta^{2}\gg1$.\\ 
 
Introducing the Compton wavelength, $\lambda_\gamma$ say, associated 
with the effective mass of the photon and defined by 
\begin{equation} 
\lambda_\gamma\equiv|\omega\Delta_\lambda|^{-1/2}, 
\end{equation} 
the required condition to be in a large radius regime can be rephrased as 
$\lambda_\gamma <R$, i.e. that the average scale associated with the 
photon is smaller than the radius of the extra--dimensions. The latter 
is expected to be of the order of the centimeter for two 
extra--dimensions.\\ 
 
In the two following sections, we derive these results in details for a
five and six dimensional spacetime. Let us stress here than when $n>3$
the oscillation probability will strongly depend on the cut--off in
energy in which case a more precise knowledge of the whole theory and
the exact experimental situation (to take into account decoherence
effect) are needed. We emphasize that in the following we have set the
cut--off to its maximum value in order to be very general. The computed
probability is thus the maximum one and the bounds on the parameters
used to derive it are the most stringent.

\section{Estimation of the probability in a five dimensional 
spacetime}\label{par3_3}  
 
In this section, we present the computation of the eigenvalues and of 
the oscillation probability when $n=1$. As it will be seen the 
computation is easier in this case because the sums in 
(\ref{f2}-\ref{y}) can be expressed in terms of circular or hyperbolic 
functions. Although the case $n=1$ is generally regarded as in 
contradiction with observation (see however e.g. \cite{randall99}), this 
computation will teach us a lot about the mixing with a large number of 
particles. 
 
 In a five dimensional world, the mixing matrix ${\cal M}$  
is explicitely given by  
\begin{equation}\label{M5D}  
  {\cal M}_\lambda= 
  \left(\begin{array}{ccccccc}  
  \Delta_\lambda & \Delta_M &\Delta_M&\Delta_M&\Delta_M&\Delta_M&\cdots\\  
  \Delta_M&0& 0&0& 0&0&\cdots\\  
  \Delta_M& 0 & \frac{-1}{2R^2\omega}&0&0&0&\cdots \\  
  \Delta_M&0&0& \frac{-1}{2R^2\omega} &0&0&\cdots\\  
  \Delta_M&0&0&0&\frac{-4}{2R^2\omega} & 0&\cdots \\  
  \Delta_M&0&0&0&0&\frac{-4}{2R^2\omega} &\ddots \\  
  \vdots&\vdots& \vdots&\vdots&\vdots&\ddots&\ddots  
  \end{array}\right).  
\end{equation}  
This matrix can be compared to the one obtained for neutrino oscillations 
is spacetime with extra--dimensions (see e.g. \cite{dudas}). 
We see on that example that for $q\not=0$ each $\Delta_m^{(q)}$  
is twice degenerated so that $r_1=1$, $r_i=2i-1$ and $s_i=2$ for $i>1$.  
  
The characteristic eigenvalues equation $\hbox{det} ({\cal M}- x  
I)=0$, in that simple case, can be obtained by developing the  
determinant of order $2N+2$ with respect to the last line to get a  
recursion relation with the determinant of order $2N$ and to find the  
limit of this series. Indeed it leads to the same result that the general  
equation (\ref{polcarfin}). 
  
With the notations of the former paragraph, the eigenvalues equation 
(\ref{polcarfin}) can now be rewritten after resummation (see 1.217 
in\cite{grad}) as 
\begin{equation}\label{eq5d}  
  y-1=\frac{\alpha^2}{\beta}{\cal K}(y/\beta),  
\end{equation}  
where the function ${\cal K}$ is defined by  
\begin{equation}\label{defK}  
{\cal K}(x)\equiv\pi|x|^{-1/2}  
\left\lbrace  
  \begin{array}{ll}  
  \cot\pi|x|^{1/2}&(x>0)\\  
  -\coth\pi|x|^{1/2}&(x<0).  
  \end{array}\right.  
\end{equation}  
The oscillation probability is then given by (\ref{ppp}) 
where the coefficients (\ref{f2}) are now reexpressed after 
resummation as 
\begin{equation}\label{y5d}  
f^2_{x_j}=\left[1+\frac{\alpha^2}{\beta^2}{\cal I}(y_j/\beta)  
\right]^{-1}  
\end{equation}  
where the function ${\cal I}\equiv\sum_{k\in Z}(x-k^2)^{-2}$ is  
obtained from (\ref{defK}) as  
\begin{equation}\label{defF}  
{\cal I}(x)\equiv\frac{\pi}{2|x|^{3/2}}  
\left\lbrace  
  \begin{array}{lc}  
   \cot\pi\sqrt{x}+\pi \sqrt{x}(1+\cot^2\pi \sqrt{x})&\qquad(x>0)\\  
   \coth\pi\sqrt{|x|}-\pi \sqrt{|x|}(1-\coth^2\pi\sqrt{|x|})&\qquad(x<0).\\  
  \end{array}\right.  
\end{equation}  
 
We now determine an approximation of the solutions of the eigenvalues 
equation (\ref{eq5d}) and then of the oscillation probability in the 
cases $\beta<0$ and $\beta>0$ assuming that $\alpha^2\ll1$. 
 
\subsection{$\beta<0$}\label{5neg} 
 
\subsubsection{Eigenvalues} 
 
\begin{figure}  
\centering  
\epsfig{figure=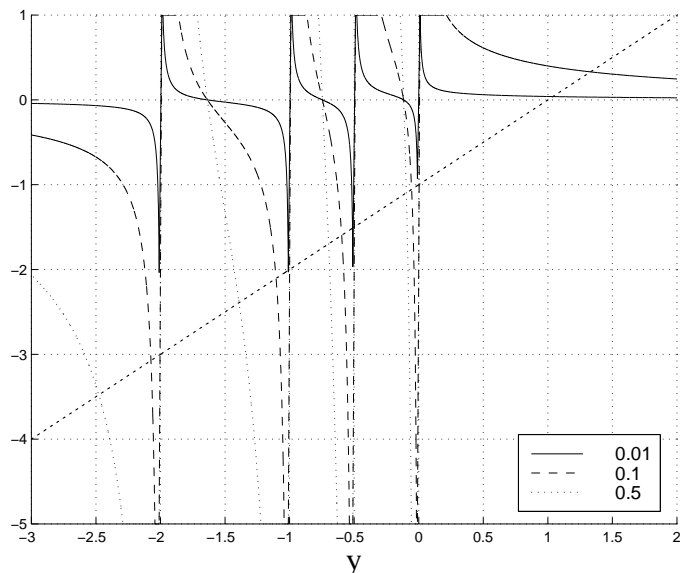,width=9cm}  
\caption{Example of a graphic resolution of equation (\ref{eq5d})  
where we have chosen $\alpha^2=(0.01,0.1,0.5)$ and $\beta=-0.5$.}  
\label{plot1}  
\end{figure}

Setting $\bar\beta\equiv-\beta>0$, we have $\beta_j= 
-(j-1)^2\bar\beta$ and one can easily sort out  
 that the solutions of equation (\ref{eq5d}) are such that 
\begin{eqnarray} 
y_1>1,\quad y_{j+1}\in\left]\beta_{j+1},\beta_{j}\right[ 
\end{eqnarray} 
with 
\begin{eqnarray} 
y_1&\quad\hbox{solution of}\quad&y-1=\pi\frac{\alpha^2}{\bar\beta} 
\sqrt{\frac{\bar\beta}{y}}\coth\pi\sqrt{\frac{y}{\bar\beta}},\label{g1}\\ 
y_{j>1}&\quad\hbox{solutions of}\quad&1-y=\pi\frac{\alpha^2}{\bar\beta} 
\sqrt{\frac{\bar\beta}{-y}}\cot\pi\sqrt{\frac{-y}{\bar\beta}}.\label{g2} 
\end{eqnarray} 
In figure \ref{plot1}, we depict the graphical resolution of this equation. 
 
The resolution of equation (\ref{eq5d}) then splits into the three 
following cases:  
 
\begin{enumerate} 
\item\underline{$y_1$}: When $y_1/\bar\beta\ll1$, (\ref{g1}) reduces 
to $y-1=\alpha^2/y$ so that $y_1\simeq 1+\alpha^2$, and one can then 
check that the condition $y_1/\bar\beta\ll1$ is equivalent to 
$\bar\beta\gg1$.\\ 
When $y_1/\bar\beta\gg1$, we set $y_1=1+\epsilon$ with $\epsilon>0$ and 
(\ref{g1}) implies that $0<\epsilon=\pi\alpha^2(1+\epsilon)^{-1/2}/ 
\bar\beta^{1/2}<\pi\alpha^2/\bar\beta^{1/2}$. Then if 
$\alpha^2/\bar\beta^{1/2}\ll1$ we deduce that $y_1\simeq1+\pi 
\alpha^2/\bar\beta^{1/2}$ and that the initial condition on $y_1$ is 
equivalent to $\bar\beta\ll1$.\\ 
In conclusion, if $\alpha^2/\bar\beta^{1/2}\ll1$, 
\begin{equation} 
y_1\simeq\left\lbrace 
\begin{array}{lc} 
1+\alpha^2&(\bar\beta\gg1)\\ 
1+\pi\frac{\alpha^2}{\bar\beta^{1/2}}&(\bar\beta\ll1). 
\end{array}\right. 
\end{equation} 
Note that these two solutions can be rewritten under the more compact 
form 
\begin{equation} 
y_1\simeq1+\frac{\alpha^2}{\beta}{\cal K}(\beta^{-1}). 
\end{equation} 
 
\item\underline{$y_{j+1}; j>1$}: Since $y_{j+1}$ satisfies 
$$(j-1)<\sqrt{\frac{-y_{j+1}}{\bar\beta}} <j$$ 
we set $\sqrt{-y_{j+1}/\bar\beta}\equiv(j-1)+\epsilon_{j+1}$ with 
$0<\epsilon_{j+1}<1$ and equation (\ref{g2}) rewrites as 
\begin{eqnarray}\label{g4} 
1+\bar\beta(j-1)^2\left[1+\frac{\epsilon_{j+1}}{j-1}\right]^2= 
\frac{\pi\alpha^2}{\bar\beta(j-1)}\frac{\cot\pi\epsilon_{j+1}} 
{1+\frac{\epsilon_{j+1}}{j-1}}. 
\end{eqnarray} 
Now, if $\alpha^2/\bar\beta\ll1$, the l.h.s. of (\ref{g4}) being 
larger than unity, it implies that $\cot\pi\epsilon_{j+1}\gg1$ which 
thus behaves as $1/\pi\epsilon_{j+1}$. We can then solve (\ref{g4}) 
for $\epsilon_{j+1}$ to get 
\begin{equation} 
y_{j}\simeq\beta_{j-1}-\frac{2\alpha^2}{1-\beta_{j-1}}. 
\end{equation} 
This expansion is valid whatever the magnitude of $\bar\beta$ as long 
as $\alpha^2/\bar\beta\ll1$.

\item\underline{$y_2$}: $y_2$ is the solution of (\ref{g2}) such that 
$0<\sqrt{-y_2/\bar\beta}<1/2$. Setting $z\equiv-y_2/\bar\beta$, 
(\ref{g2}) leads to 
\begin{eqnarray}\label{g6} 
1+\bar\beta z=\pi\frac{\alpha^2}{\bar\beta}\frac{\cot\pi\sqrt{z}}{\sqrt{z}} 
\end{eqnarray} 
with $0<\sqrt{z}<1/2$. The l.h.s. of (\ref{g6}) being greater than 
unity, it implies that, when $\alpha^2/\bar\beta\ll1$, 
$\cot(\pi\sqrt{z})/\sqrt{z}\gg1$ and thus behaves as $1/\pi\sqrt{z}$. 
At lowest order (\ref{g6}) then leads to $z\simeq\alpha^2/\bar\beta$ 
and then 
\begin{equation} 
y_2\simeq-\alpha^2. 
\end{equation} 
Again, this solution is valid whatever $\bar\beta$ such that 
$\alpha^2/\bar\beta\ll1$. 
 
\item\underline{Summary}: When $\alpha^2\ll1$, the roots of 
(\ref{eq5d}) are well approximated by 
\begin{eqnarray} 
y_1&\simeq&1+\frac{\alpha^2}{\beta}{\cal K}(\beta^{-1})\nonumber\\ 
y_{j>1}&\simeq&\beta_{j-1}-\frac{s_{j-1}\alpha^2}{1-\beta_{j-1}} 
\end{eqnarray} 
\end{enumerate} 
for all $\beta$ such that $\alpha^2\ll|\beta|$. 
 
\subsubsection{Oscillation probability} 
 
Assuming that $\alpha/|\beta|\ll1$, we can now expand (\ref{y5d}) to get 
the following behaviours of the coefficients $f_{x_i}^2$  
\begin{eqnarray} 
f^2_{x_1}&\simeq&1-\frac{\alpha^2}{\beta^2}{\cal I}(\beta^{-1}), \\ 
f^2_{x_{j>1}}&\simeq&\frac{s_{j-1}}{(1-\beta_{j-1})^2}\alpha^2.\\ 
\end{eqnarray} 
One can easily check that, at this order, $\sum f^2_{x_i}=1$. Using the 
form (\ref{ppp}) of the oscillation probability we deduce that 
\begin{equation} 
P(\gamma\rightarrow 
g)\simeq4\alpha^2\sum_{j\geq1}\frac{s_j}{(1-\beta_{j})^2} 
\sin^2\left(\frac{1-\beta_j}{2}\Delta_\lambda u\right), 
\end{equation} 
as announced in (\ref{E65}).  It can be checked that the dominant 
contribution to the probability comes from the terms 
$f^2_{x_1}f^2_{x_j}$ in (\ref{ppp}). 
 
It is worth noting that the cut--off of the theory does not enter the 
result, due to the fact that in this special case all the sums are 
converging. Since $\beta_j\leq0$ and $\alpha\ll1$, we are always in the 
weak mixing limit and the oscillation probability is well approximated 
by the sum of all the individual oscillation probabilities. The 
individual oscillation lengths are given by 
$$\ell_{\rm osc}^{(j)}=\frac{2\pi}{\Delta_\lambda}\frac{1}{1-\beta_j}< 
\frac{2\pi}{\Delta_\lambda}=\ell_{\rm osc}^{(1)}.$$ 
  
\subsection{$\beta>0$} 
 
\subsubsection{Eigenvalues} 
 
\begin{figure}  
\centering  
\epsfig{figure=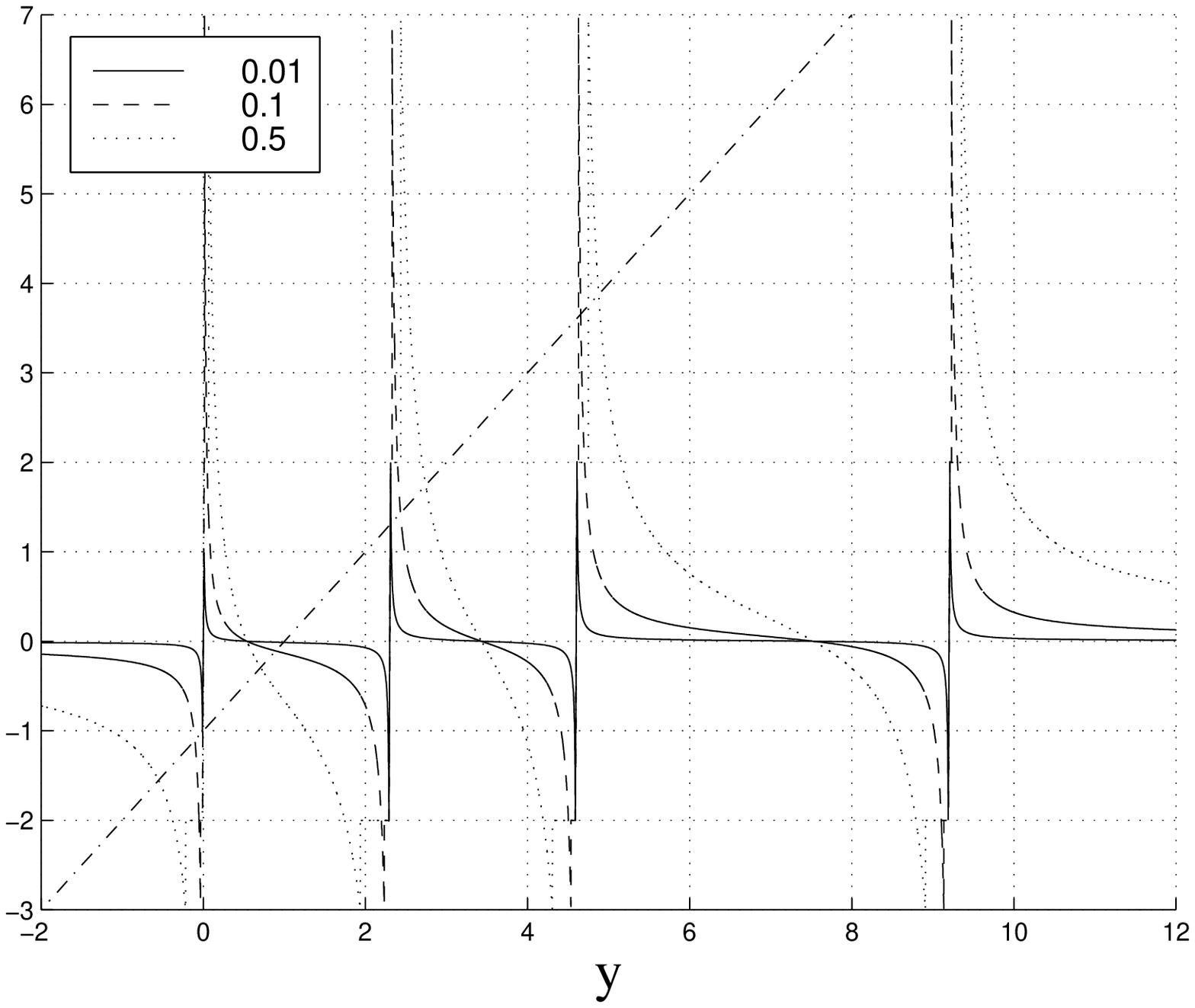,width=9cm} 
\epsfig{figure=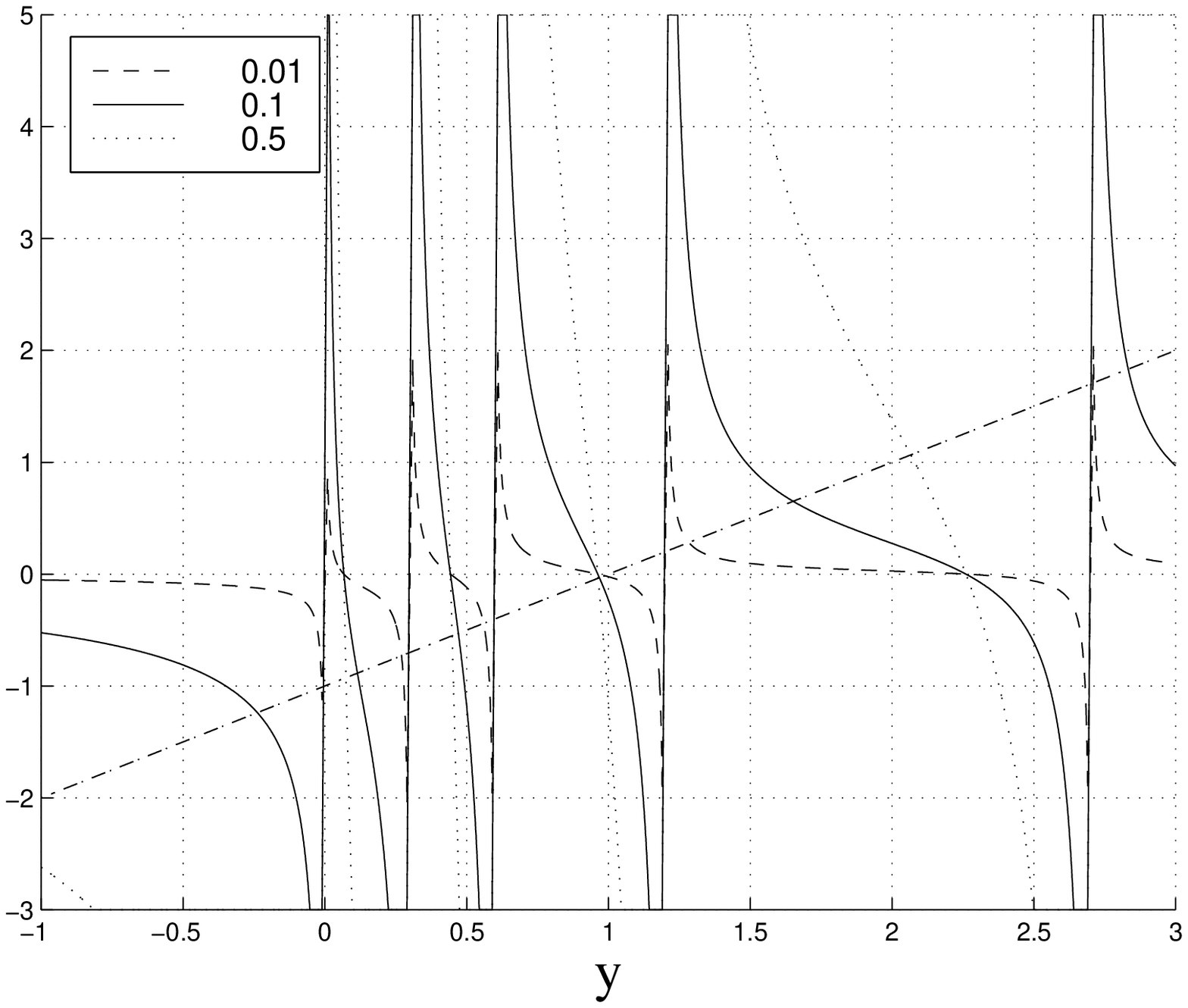,width=9cm} 
\caption{Examples of a graphic resolution of equation (\ref{y}) 
in the case $\beta>0$.  
We have chosen $\alpha^2=(0.01,0.1,0.5)$. The left 
plot describes the situation $\beta=2.3$  
and the right plot the case where $\beta=0.3$. In the latter case 
$K=4$.}  
\label{plot2}  
\end{figure}

Since $\beta_j=(j-1)^2\beta$, one can easily show that the roots 
of (\ref{eq5d}) are such that 
\begin{equation} 
y_1<0,\quad\beta_i<y_{j+1}<\beta_{j+1} 
\end{equation} 
with 
\begin{eqnarray} 
y_1&\quad\hbox{solution of}\quad&1-y=\pi\frac{\alpha^2}{\beta} 
\sqrt{\frac{\beta}{-y}}\coth\pi\sqrt{\frac{-y}{\beta}},\label{g41}\\ 
y_{j>1}&\quad\hbox{solutions of}\quad&y-1=\pi\frac{\alpha^2}{\beta} 
\sqrt{\frac{\beta}{y}}\cot\pi\sqrt{\frac{y}{\beta}}.\label{g42} 
\end{eqnarray} 
In figure \ref{plot2} we depict the graphic resolution of this equation. 
We introduce $K$ the index of the closest $\beta_i$ to unity. Contrary 
to the previous case, the discussion has to be split in four steps: 
 
\begin{enumerate} 
\item\underline{$y_1$}: When $-y_1/\beta\ll1$, (\ref{g41}) implies 
that $1-y_1\simeq-\alpha^2/y_1$ so that $y_1\simeq-\alpha^2$ and the 
initial condition reduces to $\alpha^2/\beta\ll1$. Thus when 
$\alpha^2/\beta\ll1$, whatever the magnitude of $\beta$, 
\begin{equation} 
y_1\simeq-\alpha^2. 
\end{equation} 
 
\item\underline{$y_j, 1<j<K$}: We set $\sqrt{{y_j}/{\beta}}=(j-1)-\epsilon_j$ 
with $1>\epsilon_j>0$, so that (\ref{g42}) rewrites as 
\begin{equation}\label{g47} 
1-y_j=\frac{\pi\alpha^2}{\beta}\frac{\cot\pi\epsilon_j}{(j-1)-\epsilon_j}. 
\end{equation} 
Since the l.h.s. of (\ref{g47}) is positive we deduce that $\epsilon_j<1/2$. 
Now, taking into account the fact that $1\in](\beta_K+\beta_{K-1})/2, 
(\beta_K+\beta_{K+1})/2[$, we deduce that  
$1-y_j>1-y_{K-1}>(\beta_K-\beta_{K-1})/2=\beta(K-3/2)/2$ and thus 
(since $K\geq2$) that 
$1-y_j\geq\beta/4$. 
If $\alpha^2/\beta^2\ll1$, then (\ref{g47}) implies that  
$\cot\pi\epsilon_j\simeq1/\pi\epsilon_j\ll1$ from which we deduce 
$\epsilon_j$ and then 
\begin{equation} 
y_{1<j<K}\simeq\beta(j-1)^2-\frac{2\alpha^2}{1-\beta(j-1)^2}. 
\end{equation} 
 
\item\underline{$y_j, j>K+1$}: The argument follows the same lines as 
the previous one but we now set $\sqrt{{y_j}/{\beta}}=(j-2)+\epsilon_j$ 
with $1>\epsilon_j>0$. We can now deduce from $y_j-1\geq y_{K+2}-1$ 
that $y_j-1\geq\beta/2$ and then that if $\alpha^2/\beta^2\ll1$, 
\begin{equation} 
y_{j>K+1}\simeq\beta(j-2)^2+\frac{2\alpha^2}{\beta(j-2)^2-1}. 
\end{equation} 
\item\underline{$y_K, y_{K+1}$}: If $K>1$, we set $y_K=\beta_K(1+\epsilon)$ 
and $y_{K+1}=\beta_K(1+\epsilon')$, we can use the property of 
equation (\ref{g42}) to deduce, as before, that 
\begin{equation} 
0<\epsilon<\frac{K-3/4}{(K-1)^2} 
\quad{\rm and}\quad 
0<-\epsilon'<\frac{K-5/4}{(K-1)^2} 
\end{equation} 
and then if $\beta$ is small, we can conclude that $\epsilon$ 
and $\epsilon'$ are small compared to unity. Setting $\delta\equiv 
\beta_K-1$, $\epsilon$ and $\epsilon'$ are solution of (\ref{g42})  
which reduces to 
\begin{equation} 
\delta+\epsilon\simeq\frac{2\alpha^2}{\epsilon}\quad\Longrightarrow\quad 
2\epsilon\simeq\delta\pm\sqrt{\delta^2+8\alpha^2}. 
\end{equation} 
Now, if $\delta\gg2\sqrt{2}\alpha$, we deduce that 
\begin{equation} 
y_K,y_{K+1}\in\{\beta_K+\frac{2\alpha^2}{\beta_K-1}, 
1-\frac{2\alpha^2}{\beta_K-1}\}, 
\end{equation} 
depending on the sign of $\delta$ and with the constraint 
$y_K<y_{K+1}$. 
On the other hand, if $\delta\ll2\sqrt{2}\alpha$, 
\begin{equation} 
y_K\simeq1+\alpha\sqrt{2},\quad 
y_{K+1}\simeq1-\alpha\sqrt{2}. 
\end{equation} 
\end{enumerate} 
Note that if $K=1$, then $\beta\geq2$ and the above discussion is still 
valid, but we just have the two classes of solutions  $y_1$ and 
$y_{j>1}$. 
 
\subsubsection{Oscillation probability} 
 
Assuming that $\alpha/\beta\ll1$, we can expand (\ref{y5d}) to get 
the following forms for the coefficients $f_{x_i}^2$. Assuming that 
$\beta_K-1<0$, 
\begin{eqnarray} 
f^2_{x_{1\leq j<K}}&\simeq& \frac{s_j}{(1-\beta_j)^2}\alpha^2\label{98}\\ 
f^2_{x_K}&\simeq&\left\lbrace 
                  \begin{array}{lc} 
                           \frac{1}{2}&(|\beta_K-1|\ll2\sqrt{2}\alpha)\\ 
                            \frac{s_{K}}{(1-\beta_{K})^2}\alpha^2 
                                   &(|\beta_K-1|\gg2\sqrt{2}\alpha) 
                  \end{array}\right.\label{99}\\ 
f^2_{x_{K+1}}&\simeq&\left\lbrace 
                  \begin{array}{lc} 
                           \frac{1}{2} &(|\beta_K-1|\ll2\sqrt{2}\alpha)\\ 
                            1-\frac{\alpha^2}{\beta^2}{\cal I}(\beta^{-1}) 
                                   &(|\beta_K-1|\gg2\sqrt{2}\alpha) 
                  \end{array}\right.\label{100} \\ 
f^2_{x_{j>K+1}}&\simeq&\frac{s_{j-1}}{(1-\beta_{j-1})^2}\alpha^2. \label{101} 
\end{eqnarray} 
When $\beta_K-1>0$, $f^2_{x_{K-1}}$ is then given by (\ref{100}) and 
$f^2_{x_{K+1}}$ is of the form (\ref{101}). 
 
It can be checked that, at lowest order $\sum f^2_{x_i}=1$. One can 
further check that when $|\beta_K-1|\ll2\sqrt{2}\alpha$ the first correction 
to (\ref{99}--\ref{100}) is $-\alpha^2{\cal I}(\beta^{-1})/2\beta^2$ 
and that $\sum f^2_{x_i}=1$ is also satisfied (which indeed 
has to be order by order).  Now, the 
oscillation probability (\ref{ppp}) reduces to 
\begin{equation}\label{102} 
P(\gamma\rightarrow g)\simeq 
\left\lbrace 
\begin{array}{lc} 
          (1-\epsilon) \sin^2\left[\Delta_M s_K^{1/2}u\right]
	  +4\alpha^2\sum_{i\not=K}
\frac{s_i}{(1-\beta_i)^2}\sin^2\left[\frac{1-\beta_i}{2}\Delta_\lambda u\right]
&(|\beta_K-1|\ll2\sqrt{2}\alpha) 
\\ 
           4\alpha^2\sum_{i\geq1}\frac{s_i}{(1-\beta_i)^2} 
           \sin^2\left[\frac{1-\beta_i}{2}\Delta_\lambda u\right]& 
           (|\beta_K-1|\gg2\sqrt{2}\alpha), 
\end{array}\right. 
\end{equation} 
with $\eta\equiv\alpha^2{\cal I}(\beta^{-1})/\beta^2$ 
as announced in (\ref{E65}).  Thus, when $|\beta_K-1|\gg\alpha$, we are 
in a weak mixing regime and the probability is obtained by summing over 
all the individual probabilities. Otherwise, we are in a regime of 
strong mixing with the state $K$, and the oscillation length with this state is now given by 
$$\ell_{\rm osc}=\frac{\pi}{\Delta_M s_K^{1/2}}.$$ 
  
\subsection{Summary and discussion} 
 
In the limit where $\alpha^2\ll1$, we have estimated the oscillation 
probability (\ref{ppp}) to be 
\begin{itemize} 
\item\underline{$\beta<0$}:  
$$P(\gamma\rightarrow g)\simeq4\alpha^2\sum_{i\geq1}\frac{s_i}{(1-\beta_i)^2} 
           \sin^2\left[\frac{1-\beta_i}{2}\Delta_\lambda u\right],$$ 
 
\item\underline{$\beta>0$}: 
$$P(\gamma\rightarrow g)\simeq 
\left\lbrace 
\begin{array}{lc} 
I \sin^2\left[\Delta_M s_K^{1/2}u\right]+4\alpha^2\sum_{i\not=K} 
\frac{s_i}{(1-\beta_i)^2}\sin^2\left[\frac{1-\beta_i}{2}\Delta_\lambda 
u\right]&(|\beta_K-1|\ll2\sqrt{2}\alpha)\\ 
           4\alpha^2\sum_{i\geq1}\frac{s_i}{(1-\beta_i)^2} 
           \sin^2\left[\frac{1-\beta_i}{2}\Delta_\lambda u\right]& 
           (|\beta_K-1|\gg2\sqrt{2}\alpha^2), 
\end{array}\right.$$ 
\end{itemize} 
with $I\equiv1-\alpha^2{\cal I}(\beta^{-1})/\beta^2$, 
as announced in (\ref{E63}) and (\ref{E65}).  When 
$|\beta_j-1|\gg2\sqrt{2}\alpha^2$ for all $j$, it is interesting to study the 
limit where $u\gg\ell_{\rm osc}$ in which case we can assume that the 
sines can be replaced by their average value to get the two limiting 
behaviours: 
\begin{eqnarray} 
P(\gamma\rightarrow 
g)&\simeq&2\alpha^2\left(1+\frac{\pi^4}{45\beta^2}\right) 
\qquad (|\beta|\gg1),\label{5dfin1}\\ 
P(\gamma\rightarrow 
g)&\simeq&\pi\frac{\alpha^2}{\sqrt{|\beta|}} 
\quad\qquad (|\beta|\ll1).\label{5dfin2} 
\end{eqnarray} 
As explained in \S~\ref{par3}, we respectively see on 
(\ref{5dfin1}-\ref{5dfin2}) the {\it small} and {\it large} radius 
regimes where the extra--dimensions either have no effect (\ref{5dfin1}) 
or enhance (\ref{5dfin2}) the probability. We also find the regime of 
strong mixing with the $K^{\rm th}$ KK graviton and the effect on the 
oscillation length as discussed below equation (\ref{E64}).

\section{Estimation of the probability in a six dimensional spacetime}  
\label{DD}  
 
Let us now turn to the physically more interesting case of $n=2$ 
extra--dimensions. Now, equation (\ref{y}) cannot be solved exactly in 
general, but its solutions can be well approximated when the coupling 
between the photon and the graviton (this coupling is measured by 
$\Delta_M$ in the matrix (\ref{M})) is small enough compared to the 
typical mass parameters of the mixing particles [i.e. the diagonal terms 
in (\ref{M})].  We follow the same lines as in the previous section.

\subsection{$1 \geq \beta>0$}  
 
\subsubsection{Eigenvalues} 
 
In that case, the plasma effects dominate over the vacuum polarisation 
in $\Delta_\lambda$; all the $\beta_i$ are positive and one can easily  
show that the roots of (\ref{y}) are such that 
\begin{eqnarray}\label{intey} 
y_1<0,\quad y_i\in ]\beta_{i-1},\beta_i[, 
\quad y_{N_D+1}>\beta_{N_D}. 
\end{eqnarray}

We introduce $k(i)$ the index such that $\beta_{k(i)}$ is the closest 
$\beta_i$ to $y_i$. From (\ref{intey}) one has that $k(i)\in\{i-1,i\}$.  
The eigenvalue equation (\ref{y}) for the root $y_i$ can be rewritten as 
\begin{equation}\label{yi1} 
y_i-1=\alpha^2\frac{s_k}{y_i-\beta_k}+{\cal F}_k(y_i), 
\end{equation} 
where ${\cal F}_{k}$ is defined as 
\begin{equation} 
 \label{deff} 
{\cal F}_{k}(y)\equiv\alpha^2\sum_{j\not=k} 
\frac{s_j}{y-\beta_j}. 
\end{equation} 
${\cal F}$ with no subscript denotes the function defined by the sum  
(\ref{deff}) taken over all indices $j$ from $j=1$ to $j=N_D$. 
To finish, we introduce the index $K$ such that $\beta_K$  
is the closest $\beta_i$ to unity and then, 
\begin{equation} 
\forall i\not= K \quad |\beta_i-1| \geq \frac{\beta}{2}. 
\end{equation} 
 
As in the five dimensional case, the determination of the roots $y_i$  
has to be split in the three following cases: 
\begin{itemize} 
\item \underline{$i\leq K-1$}:  
We first show that $k(i)=i$. 
For that purpose, we consider the function ${\cal H}(y)$ defined by  
\begin{equation} 
{\cal H}(y) \equiv 1-y+{\cal F}(y). 
\end{equation} 
This function is strictly decreasing on $]\beta_{i-1},\beta_i[$  
(${\cal H}$   vanishes only once in this interval in $y=y_i$). 
Showing that ${\cal H}\left(\frac {\beta_i + \beta_{i-1}}{2}\right) \geq 0$, is then 
 enough to prove that $k(i)=i$. Since one has 
\begin{equation} \label{E108} 
1 -\frac {\beta_i + \beta_{i-1}}{2} \geq \frac{\beta}{2}   
\quad\hbox{ and }\quad\forall k\quad 
\left|\frac{\beta_i + \beta_{i-1}}{2} - \beta_k \right| \geq \frac{\beta}{2},  
\end{equation} 
using  (\ref{C22}), one obtains  
\begin{equation} \label{E109} 
\left|{\cal F} \left( \frac{\beta_i + \beta_{i-1}}{2} \right)\right| 
 \leq {\cal Q}  \frac{\alpha^2}{\beta^2} \hbox{sup} 
\left({\cal Q'},\sqrt{\beta y}\right)\leq{\cal 
 Q} {\cal Q'}\frac{\alpha^2}{\beta^2} \quad  
\hbox{for $y\leq 1$ and $\beta \leq 1$}. 
\end{equation} 
(the constants ${\cal Q}$ and ${\cal Q'}$ are defined in equation (\ref{C22})  
of appendix \ref{appD}). 
Comparing (\ref{E108}) and (\ref{E109}) one sees that for $\alpha$ smaller enough than   
$\beta$ (namely $2{\cal Q}{\cal Q'}\frac{\alpha^2}{\beta^3}<1)$,  
one has  ${\cal H} \left(\frac {\beta_i + \beta_{i-1}}{2} 
\right) >0$ and then that $k(i)=i$. 
 
We will assume in the following the slightly stronger constraint 
\begin{equation} 
10 {\cal Q}{\cal Q'}\frac{\alpha^2}{\beta^3}<1. \label{cont1} 
\end{equation} 
Now, we set $y_i=\beta_i-\epsilon_i$, 
with $\epsilon_i>0$. Equation (\ref{yi1}) can be rewritten as an 
equation for $\epsilon_i$, with ${\cal F}_i\equiv{\cal F}_i(y_i)$, 
\begin{equation}  
\frac{\epsilon_i^2}{\beta_i-1-{\cal F}_i}- \epsilon_i -\frac{\alpha^2s_i} 
{\beta_i-1-{\cal F}_i} =0, 
\end{equation} 
the positive solution of which is  given to leading order by 
\begin{equation} 
\epsilon_i \simeq \frac{\alpha^2s_i}{1-\beta_i}, \label{E113} 
\end{equation} 
when  
$\alpha$ and  
$\beta$ verify the constraint (\ref{cont1}) 
\footnote{to establish this we have used (\ref{C22}) and (\ref{C26}) and 
showed that (\ref{cont1}) leads to $|\beta_i -1| \gg {\cal F}_i$ and 
$|\beta_i -1| \gg 2 \alpha \sqrt{s_i}$ which in turn leads to the expression 
(\ref{E113}). In the rest of this section $a \gg b$ means that $a > 10 
b$ which we assumed to be enough to neglect b with respect a.}.  The 
eigenvalues are then given at leading order by 
\begin{equation} \label{appik1} 
y_i \simeq \beta_i-\alpha^2\frac{s_i}{1-\beta_i}. 
\end{equation} 
 
\item \underline{$i\geq K+2$}: 
Using a similar line of reasoning as in the previous case, one can show that $k(i)=i-1$  
and then that $y_i$ is 
given at dominant order by (for $\alpha$ and $\beta$ verifying (\ref{cont1}))   
\begin{equation} 
y_i \simeq \beta_{i-1}+\alpha^2\frac{s_{i-1}}{\beta_{i-1}-1}. 
\end{equation} 
 
\item \underline{$i=K,K+1$}: We first estimate the root $y_K$.  We 
assume that $1 \in [\beta_K, \beta_{K+1}[$ (similar conclusions can be 
obtained when $1 \in [\beta_{K-1}, \beta_K[$), we have 
\begin {equation} 
1-\frac{\beta_K + \beta_{K-1}}{2} > \frac{\beta}{2}. 
\end{equation} 
As in the previous case, this is enough to show that $y_K \in 
[\beta_{K-1}+ \frac{\beta_K - \beta_{K-1}}{2}, \beta_{K}[$ 
\footnote{here we use (\ref{cont1}) again.}.  We set $y_K=\beta_K-\epsilon_K$ 
and ${\cal F}_K\equiv{\cal F}_K(y_K)$ with $\epsilon_K>0$ solution of 
\begin{equation} \label{epsiepsi} 
\frac{\epsilon_K^2}{\beta_K-1-{\cal F}_K}- \epsilon_K -\frac{\alpha^2s_K} 
{\beta_K-1-{\cal F}_K} =0, 
\end{equation} 
the positive root of which is 
\begin{equation}  
\epsilon_K=\frac {1-\beta_K+{\cal F}_K}{2}\left(-1 +\sqrt{1+\frac{4\alpha^2s_K} 
{(\beta_K-1-{\cal F}_K)^2}} \right). 
\end{equation} 
For $\alpha$ smaller enough than  $\beta $  
\footnote{One has here to impose a slightly stronger condition 
than the previous one: namely $4{\cal Q}{\cal Q'}\frac{\alpha}{\beta^2}<1$ in order to be able to neglect  
${\cal F}_K$ with respect to $2\alpha \sqrt{s_K}$. This also insures that the expressions (\ref{E120}) are 
valid.  \label{foot}}, one can consider the two limiting regimes (we will not 
consider here the intermediate case, in order to simplify the discussion) 
\begin{equation} \label{E120} 
\epsilon_{K}\simeq\left\lbrace 
\begin{array}{lcl} 
\alpha \sqrt{s_K}&\hbox{if}& |\beta_K-1| \ll 2 \alpha \sqrt{s_K} \\ 
\frac{\alpha^2 s_K}{1-\beta_K}& 
\hbox{if}& |\beta_K-1| \gg 2 \alpha \sqrt{s_K}. 
\end{array}\right. 
\end{equation} 
Let us now turn to the evaluation of the root $y_{K+1}$.  The discussion 
mimics the previous one. Assuming that 1 is not too close to 
$\frac{\beta_K+\beta_{K+1}}{2}$ \footnote{namely 
$\left|1-\frac{\beta_K+\beta_{K+1}}{2} \right| > \frac{\beta}{10}$. When 
this is not the case, one obtains the same results as in (\ref{E123}) 
when $|\beta_K-1| \gg 2 \alpha \sqrt{s_K}$.}, one can show that $y_{K+1} 
\in ]\beta_K,\beta_K+\frac{\beta_{K+1}-\beta_{K}}{2}[$ under the 
condition (\ref{cont1}).  Then we write  
$y_{K+1}=\beta_K-\epsilon_{K+1}$, with $\epsilon_{K+1} < 0$. 
$\epsilon_{K+1}$ is solution of equation (\ref{epsiepsi}) with 
${\cal F}_{K}\equiv{\cal F}_{K}(y_{K+1})$. One considers (under the condition 
on $\alpha$ and $\beta$ of footnote \ref{foot}) the two limiting regimes 
\begin{equation} \label{E123} 
y_{K+1}\simeq\left\lbrace 
\begin{array}{lcl} 
\beta_K+\alpha\sqrt{s_K}&\hbox{for}& |\beta_K-1|\ll2\alpha\sqrt{s_K}, \\ 
1+{\cal F}(1)&\hbox{for}& |\beta_K-1|\gg2\alpha\sqrt{s_K}. 
\end{array} 
\right. 
\end{equation} 
\end{itemize} 
 
\subsubsection{Oscillation probability} 
 
We now need to expand the coefficients (\ref{f2}) in order to  
estimate the probability (\ref{ppp}).

\begin{itemize} 
\item \underline {For $i \leq K-1$}, we have from equation (\ref{f2}) and (\ref{appik1}) 
\begin{equation}  
f_{x_i}^2=\left[1+\frac{(1-\beta_i)^2}{\alpha^2 s_i} +{\cal G}_i \right]^{-1} 
\end{equation} 
with  
\begin{equation} 
{\cal G}_i\equiv {\cal G}_i(y_i) \equiv \alpha^2\sum_{k=1, k\neq i}^{N_D}\frac{s_k} 
{\left(y_i-\beta_k\right)^2}. 
\end{equation} 
Using (\ref{C24}), one can then show that under the condition (\ref{cont1}) 
\begin{equation} 
\frac{(1-\beta_i)^2}{\alpha^2 s_i} \gg {\rm max}(1,{\cal G}_i) 
\end{equation} 
so that, at dominant order, 
\begin{equation} 
f_{x_i}^2\simeq\alpha^2 \frac{s_i}{(1-\beta_i)^2}. 
\end{equation} 
\item \underline{For $i \geq K+2$}, we find in a similar way (and under the same condition) 
\begin{equation} 
f_{x_i}^2\simeq\alpha^2 \frac{ s_{i-1}}{(1-\beta_{i-1})^2}. 
\end{equation} 
 
\item \underline{For $i =K,K+1$}, we distinguish the two regimes  
$|\beta_K-1| \ll 2\alpha\sqrt{s_K}$ and $|\beta_K-1|\gg2\alpha\sqrt{s_K}$. 
 
Assuming that $|\beta_K-1|\gg2\alpha\sqrt{s_K}$, we find \footnote{Using 
(\ref{C24}), we show that this expansion is valid under the condition 
of the footnote (\ref{foot}).} the dominant contribution to $f_{x_{K}}$ 
and $f_{x_{K+1}}$ to be 
 \begin{equation} 
f_{x_{K}}^2 \simeq \alpha^2\frac{ s_K}{(1-\beta_K)^2} \qquad\hbox{and} 
\qquad f_{x_{K+1}}^2 \simeq  1. 
 \end{equation} 
When $|\beta_K-1|\ll2\alpha\sqrt{s_K}$, the dominant 
contribution to $f_{x_{K}}^2$ and $f_{x_{K+1}}^2$ are 
\begin{equation}\label{132} 
f_{x_{K}}^2 \simeq\frac{1}{2}\qquad \hbox{and}\qquad  
f_{x_{K+1}}^2 \simeq\frac{1}{2}. 
\end{equation} 
\end{itemize} 
Now, inserting these results in (\ref{ppp}), we find that the 
oscillation probability is given to dominant order, for $|\beta_K-1| 
\gg 2\alpha\sqrt{s_K}$, by 
\begin{equation} 
P(\gamma\rightarrow g)\simeq4\sum_{i\neq K+1}  
f^2_{x_i}\sin^2\left[ 
\frac{x_i-x_{K+1}}{2}u\right]\simeq 4\sum_{i=1}^{i=N_D}  
\alpha^2 \frac{ s_{i}}{(1-\beta_i)^2}\sin^2\left[ 
\frac{1-\beta_i}{2}\Delta_{\lambda} u\right] 
\end{equation} 
  
This expression, as announced, is analogous to  (\ref{E65})  
and corresponds  to the case when no KK state 
mixes strongly with the photon.

Now, when $|\beta_K-1| \ll 2\alpha\sqrt{s_K}$, one has  
\begin{eqnarray} 
P(\gamma\rightarrow g) &\simeq& 
(1-\eta)\sin^2\left[\frac{x_{K}-x_{K+1}}{2}u 
\right] 
+ 2\sum_{P=K,K+1}\sum_{i\neq P}f^2_{x_i} 
\sin^2\left[\frac{x_i-x_P}{2}u\right]\label{II} 
\end{eqnarray} 
which can be rewritten as 
\begin{equation}\label{132} 
P(\gamma\rightarrow g) \simeq (1-\eta)\sin^2 [\Delta_M \sqrt{s_K} u] +  
4\sum_{i \neq K}  
\alpha^2 \frac{ s_{i}}{(1-\beta_i)^2}\sin^2\left[ 
\frac{(1-\beta_i)}{2} \Delta_{\lambda} u\right]
\end{equation}
with $\eta\equiv\sum_{i\neq K}\alpha^2s_i/(1-\beta_i)^2\ll1$.  As in the
five dimensional case it was obtained by imposing that $\sum
f^2_{x_i}=1$ order by order.  This corresponds to the case where one KK
state mixes strongly with the photon and again the oscillation
probability is found to be equivalent to (\ref{E63}).

\subsection{$-1\leq \beta <0$}\label{parbetapos}  
  
We now consider the case where $\beta<0$ (i.e. when the vacuum  
contribution dominates over the plasma effects in $\Delta_\lambda$).   
It is easy to see graphically (see figure \ref{plot1}) that the $N_D+1$  
solutions of the eigenvalue equation (\ref{y})  are such that  
\begin{equation}  
y_1>1,\quad y_{i+1}\in\left]\beta_{i+1},\beta_i\right[,  
\quad y_{N_D+1}<\beta_{N_D}.  
\end{equation}  
 
\subsubsection{Eigenvalues} 
 
We do not detail the computation of the eigenvalues 
since it is similar to the former case. It is even 
simpler since now we do not have to single out the 
mode $K$ (look for instance to the five dimensional 
case \S~\ref{5neg}). 
 
We  have only to assume the less stringent constraint than (\ref{cont1}), $10 
\frac{\alpha^2}{\beta^2}{\cal Q}{\cal Q'}<1$, 
in order for the following expansions to be valid. Under this 
condition one can show that 
\begin{eqnarray} 
y_1&\simeq& 1+\alpha^2\sum_{i}\frac{s_i}{1-\beta_i},\nonumber\\ 
y_{j>1}&\simeq&\beta_{j-1}-\frac{s_{j-1}}{1-\beta_{j-1}}\alpha^2. 
\end{eqnarray} 
 
\subsubsection{Oscillation Probability} 
 
The coefficients (\ref{f2}) are then given at leading order by 
\begin{eqnarray} 
f^2_{x_1}&\simeq&1,\nonumber\\ 
f^2_{x_{j>1}}&\simeq&\alpha^2\frac{s_{j-1}}{(1-\beta_{j-1})^2}. 
\end{eqnarray} 
The oscillation probability (\ref{ppp}) can be expanded as 
\begin{equation} 
P(\gamma\rightarrow g)\simeq4f^2_{x_1}\sum_{j>1}f^2_{x_{j}} 
\sin^2\left[\frac{x_1-x_j}{2} u\right] 
\end{equation} 
where we have neglected higher order terms. 
This can be rewritten as 
\begin{equation} 
P(\gamma\rightarrow g)\simeq4\alpha^2 
\sum_{j}\frac{s_j}{(1-\beta_j)^2} 
\sin^2\left[\frac{1-\beta_j}{2}\Delta_\lambda u\right]\label{ph} 
\end{equation} 
which is again analogous to (\ref{E65}). 
 
 
\subsection{$|\beta|>1$} 
 
We introduce the new parameters $\bar\alpha\equiv\Delta_M/\Delta_m<0$ and 
$\bar\beta_i\equiv \Delta_m^{(r_i)}/\Delta_m=\vec{p}_{(r_i)}^2$, 
$\bar{\gamma} \equiv \Delta_\lambda/\Delta_m$ and $z\equiv x/\Delta_m$. 
The eigenvalue equation (\ref{eqxi}) can be rewritten as 
\begin{equation} 
z-\bar{\gamma}=\bar\alpha^2\sum_{i=1}^{N_D}\frac{s_i}{(z-\bar\beta_i)}= 
{\bar\alpha^2 \over z} 
+\bar\alpha^2\sum_{i=2}^{N_D}\frac{s_i}{(z-\bar\beta_i)}. 
\label{zerozero} 
\end{equation} 
One sees easily graphically that this equation admits one negative root 
$z_1$ and that the other $z_i$ ($i \geq 3$) verify $z_i \in 
]\bar{\beta}_{i-1},\bar{\beta}_i[$.  We discuss here only the case where 
$\bar{\gamma}$ is closer to $0$ than to $1$ (we assume that 
$\bar{\gamma} <{1}/{4})$.  The discussion is very similar to the 
previous cases. Under the condition $10\bar{\alpha}^2{\cal Q}{\cal 
Q'}<1$, one finds that the roots $z_i$ with $i \geq 2$ are given by 
\begin{equation} 
z_i \simeq \bar{\beta}_{i-1} + \frac{\bar{\alpha}^2 
 s_{i-1}}{\bar{\beta}_{i-1}-\bar{\gamma}}\qquad 
 \hbox{so that}\qquad 
f^2_{x_i} \simeq \frac{\bar{\alpha}^2 s_{i-1}}{(\bar{\beta}_{i-1} 
-\bar{\gamma})^2}. 
\end{equation} 
Under the condition $4 \bar{\alpha} 
{\cal Q}{\cal Q'} < 1$, the two roots $z_1$ and $z_2$ are given, for 
$\bar\gamma>0$ by\footnote{For $\bar\gamma<0$ the results are similar; 
one has only to exchange the expressions of $z_1$ and $z_2$ in the case 
$|\bar\gamma|\gg2|\bar\alpha|$.} 
by 
\begin{equation} 
 \left\lbrace 
 \begin{array}{l}  
    z_1 \simeq \bar{\alpha}\\ 
    z_2 \simeq -\bar{\alpha} 
 \end{array}\right. 
\quad\hbox{if}\quad\bar{\gamma} \ll 2|\bar{\alpha}| 
\quad\hbox{and by}\quad 
\left\lbrace 
\begin{array}{l} 
  z_1 \simeq -\frac {\bar{\alpha}^2}{\bar{\gamma}}\\ 
  z_2  \simeq \bar{\gamma} +{\cal F}(\bar{\gamma}) 
\end{array}\right. 
\quad\hbox{if}\quad\bar{\gamma} \gg 2|\bar{\alpha}| 
\end{equation} 
from which we deduce that the coefficients are given either by 
\begin{equation} 
\left\lbrace 
\begin{array}{l}  
  f^2_{x_1} \simeq \frac{1}{2}\\ 
  f^2_{x_2} \simeq \frac{1}{2} 
\end{array}\right. 
\quad\hbox{if}\quad\bar{\gamma} \ll 2|\bar{\alpha}| 
\quad\hbox{or by}\quad 
\left\lbrace 
\begin{array}{l} 
  f^2_{x_1} \simeq 1\\ 
  f^2_{x_2} \simeq \frac{\bar{\alpha}^2}{\bar{\gamma}^2} 
\end{array}\right. 
\quad\hbox{if}\quad\bar{\gamma} \gg 2|\bar{\alpha}|. 
\end{equation} 
 
When $\bar{\gamma} \ll 2 |\bar{\alpha}|$, the oscillation probability is 
given by 
\begin{equation}\label{IIbis} 
P(\gamma\rightarrow g)\simeq (1-\tilde\eta)\sin^2(\Delta_M u) + 4 
\sum_{i \geq 2}\frac{\alpha^2 s_i}{\beta_i^2} 
\sin^2\left[\frac{\beta_i}{2}\Delta_\lambda u\right] 
\end{equation} 
and when $\bar{\gamma} \gg 2 |\bar{\alpha}|$ by  
\begin{equation} 
P(\gamma\rightarrow g)\simeq  4 
\sum_{i \geq 1}\frac{\alpha^2 s_i}{(1-\beta_i)^2} 
\sin^2\left[\frac{1-\beta_i}{2}\Delta_\lambda u\right]. 
\end{equation} 
The small coefficient
$\tilde\eta\equiv\sum_{i\geq2}s_i\alpha^2/\beta_i^2$ is obtained as
in (\ref{II}).
 
\subsection{Summary}  
 
In the limit where $\alpha^2 <1$ we have estimated the oscillation 
probability (\ref{ppp}) for a six dimensional spacetime to be 
\begin{itemize} 
\item \underline{$1 \geq \beta>0$}: 
For $|\beta_K-1| \gg 2\alpha\sqrt{s_K}$, 
\begin{equation} 
P(\gamma\rightarrow g)\simeq 4\sum_{i=1}^{i=N_D}  
\alpha^2 \frac{ s_{i}}{(1-\beta_i)^2}\sin^2\left[ 
\frac{1-\beta_i}{2}\Delta_{\lambda} u\right] 
\end{equation} 
and for $|\beta_K-1| \ll 2\alpha\sqrt{s_K}$ 
\begin{equation} 
P(\gamma\rightarrow g) \simeq (1-\eta)\sin^2 [\Delta_M \sqrt{s_K} u] +  
4\sum_{i \neq K}  
\alpha^2 \frac{ s_{i}}{(1-\beta_i)^2}\sin^2\left[ 
\frac{(1-\beta_i}{2} \Delta_{\lambda} u\right], 
\end{equation} 
these results being valid as long as $4{\cal Q}{\cal 
Q'}\frac{\alpha}{\beta^2}<1$ and the coefficient $\eta$ being
defined in equation (\ref{132}). 
 
\item \underline{$-1\leq \beta <0$}: 
\begin{equation} 
P(\gamma\rightarrow g)\simeq4\alpha^2 
\sum_{j}\frac{s_j}{(1-\beta_j)^2} 
\sin^2\left[\frac{1-\beta_j}{2}\Delta_\lambda u\right], 
\end{equation} 
valid if $10\frac{\alpha^2}{\beta^2}{\cal Q}{\cal Q'} < 1$. 
 
\item \underline{$|\beta|  > 1$}: 
\begin{eqnarray} 
P(\gamma\rightarrow g)&\simeq& (1-\tilde\eta)\sin^2(\Delta_M u) + 4 
\sum_{i \geq 2}\frac{\alpha^2 s_i}{\beta_i^2} 
\sin^2\left[\frac{\beta_i}{2}\Delta_\lambda u\right]\nonumber\\ 
&\simeq& 4 
\sum_{i \geq 1}\frac{\alpha^2 s_i}{(1-\beta_i)^2} 
\sin^2\left[\frac{1-\beta_i}{2}\Delta_\lambda u\right] 
\end{eqnarray} 
respectively for $\bar{\gamma} \ll 2 |\bar{\alpha}|$ and for
$\bar{\gamma} \gg 2 |\bar{\alpha}|$, the result being valid if $10
\bar{\alpha}^2 {\cal Q}{\cal Q'} < 1$ and the small coefficient
$\tilde\eta$ is defined in (\ref{IIbis}).
\end{itemize} 
 
\section{Mixing in an inhomogeneous field}\label{par4}  
  
In all the previous sections, we have assumed that the magnetic  
field was homogeneous. This is however a very crude approximation  
for most of the realistic physical systems. In this section we first  
extand our analysis to inhomogeneous magnetic fields and  
give some implications of the inhomogeneity of the external field.

\subsection{Computation of the oscillation probability}  
   
Following \cite{raffelt88}, we rewrite the equation of evolution (\ref{systfinal}) as 
a Schr\"odinger equation  
\begin{equation}\label{heisen1}  
  i\partial_u \vec {\cal V}=\left({\cal H}_0 + {\cal H}_1\right)  
  \vec {\cal V},  
\end{equation}  
where we set $\vec{\cal V} \equiv(A,G^{(0)},\ldots,G^{(N)})$. The two  
matrices ${\cal H}_0$ and ${\cal H}_1$ are respectively defined by  
\begin{equation}\label{H0}  
  {\cal H}_0(u)\equiv\omega+  
  \left(\begin{array}{ccccc}  
  \Delta_\lambda&&&&\\  
  &0&&&\\  
  &&\Delta_m^{(1)}&&\\  
  &&&\ddots&\\  
  &&&&\Delta_m^{(N)}  
  \end{array}\right)  
\end{equation}  
and  
\begin{equation}\label{H1}  
  {\cal H}_1(u)\equiv  
  \left(\begin{array}{cccc}  
  0&\Delta_M&\cdots&\Delta_M\\  
  \Delta_M&&&\\  
  \vdots&&&\\  
  \Delta_M&&&  
  \end{array}\right).  
\end{equation}  
We assume that ${\cal H}_1$ is a perturbation compared to ${\cal 
H}_0$. This approximation is equivalent to saying that 
$\Delta_M/\Delta_\lambda$ and $\Delta_M/\Delta_m$ are small compared to 
unity, i.e. that $\alpha\ll1$ and $\alpha/\beta\ll1$. When $H_0$ is 
inhomogeneous only $\Delta_M$ and $\Delta_\lambda$ depend on $u$ while 
all the $\Delta_m^{(q)}$ are constant. 
   
We first solve (\ref{heisen1}) at zeroth order, i.e. by neglecting  
${\cal H}_1$ with respect to ${\cal H}_0$, as  
\begin{equation}\label{a0}  
  \vec {\cal V}^{(0)}(u)=U(u)\vec{\cal V} ^{(0)}(0),  
\end{equation}  
where the evolution operator $U$ is defined by  
\begin{equation}\label{defu}  
  U(u)\equiv \exp{-i\int_0^u{\cal H}_0(u')du'}.  
\end{equation}  
Note that at this order there is no mixing effect since ${\cal  
H}_0$ is diagonal. 
   
The general solution of (\ref{heisen1}) is obtained by shifting to the 
``interaction representation'' where $\vec{\cal V}_{\rm int}\equiv 
U^\dag\vec{\cal V}$ so that (\ref{heisen1}) can be rewritten as 
\begin{equation}\label{heisen2}  
  i\partial_u\vec {\cal V}_{\rm int}={\cal H}_{\rm int}\vec {\cal V}_{\rm int}  
\end{equation}  
with ${\cal H}_{\rm int}\equiv U^\dag{\cal H}_1U$. This equation  
can be solved iteratively by setting $\vec{\cal V}_{\rm int}=\sum 
\vec{\cal V}_{\rm int}^{(k)}$ with 
\begin{equation}\label{solA}  
  \vec {\cal V}_{\rm int}^{(n+1)}=-i\int_0^u du' {\cal H}_{\rm int}(u')  
  \vec {\cal V}_{\rm int}^{(n)}(u'),  
\end{equation}  
with the initial condition $\vec {\cal V}_{\rm int}^{(0)}\equiv\vec {\cal V}  
(0)$.  
 
Introducing the basis $\lbrace\vec{\cal A},\vec{\cal G}_q\rbrace$ with 
$\vec{\cal A}\equiv(1,0,\ldots,0)$ and $\vec{\cal G}_q$ being 
the state of the $q^{\rm th}$ graviton, ${\cal G}_q^i=\delta^i_q$ 
for $i\in\lbrace1,\ldots,N+2\rbrace$ and starting with an initial 
state describing a pure photon, i.e. $\vec{\cal V}(0)= 
A(0)\vec{\cal A}$, we obtain 
\begin{equation} 
\vec{\cal V}_{\rm int}^{(1)}(u)=-i\int_0^udz\,\Delta_M(z) 
\sum_q\hbox{e}^{i\int_0^z(\Delta_m^{(q)}-\Delta_\lambda(y))dy} 
A(0)\vec{\cal G}_q,\label{V1} 
\end{equation} 
where we have used that ${\cal H}_0\vec{\cal A}=\Delta_\lambda\vec{\cal 
A}$, ${\cal H}_1\vec{\cal A}=\Delta_M\sum_q\vec{\cal G}_q$ and 
${\cal H}_1\vec{\cal G}_q=\Delta_m^{(q)}\vec{\cal G}_q$.  
If we restrict to the first iteration,  the oscillation probability is 
then given by 
\begin{equation}  
P(\gamma\rightarrow g)=\sum_q\left|\left<\vec{\cal G}_q\right|\left.  
\vec{\cal V}^{(0)}+\vec{\cal V}^{(1)}\right>\right|^2 
= \sum_q\left|\left<\vec{\cal G}_q\right|\left.\vec{\cal V}_{\rm 
int}^{(1)}\right>\right|^2, 
\end{equation}  
that is, using (\ref{V1}), 
\begin{equation}\label{124}  
P(\gamma\rightarrow g)=\sum_q\left|\int_0^{u}  
\Delta_M(u')\hbox{e}^{i\Delta^{(q)}_mu'-i\int_0^{u'}\Delta_\lambda  
(u'')du''}du'\right|^2.  
\end{equation}  
We can check that in a homogeneous field we recover (\ref{E65}), 
i.e. the oscillation probability in the weak mixing case.  Note that in 
the particular case of the weak mixing this method of computing the 
oscillation probability is shorter than the one used in the two former 
sections since it does not involve the determination of the eigenvalues of 
${\cal M}$. But, one has to assume that the probability is small 
compared to unity \cite{raffelt88}, which is not necessarily the case 
for instance when we are in the strong mixing regime.

\subsection{Example of applications}\label{7b} 
 
As an example, we consider the mixing in a periodic magnetic field of 
the form $H_0\cos\Delta_0u$ with $\Delta_0>0$ for which the 
oscillation probability, in the weak mixing regime, is given by 
(\ref{124}) 
\begin{equation} 
P(\gamma\rightarrow g)=\sum_{i\geq1}s_i\left|\int^u_0 dz\, 
\Delta_M\cos(\Delta_0z)\, \hbox{e}^{i\Delta_m^{(r_i)}z} 
 \hbox{e}^{-i\int_0^z\Delta_\lambda(v)dv}\right|^2. 
\end{equation} 
Assume that $\Delta_{\rm plasma}$ dominates so that we can neglect the 
variation of $\Delta_\lambda$ with $z$ 
then, the probability becomes 
\begin{equation} 
P(\gamma\rightarrow g)\simeq\Delta_M^2\sum_is_i\left( 
\frac{1}{(\Delta_m^{(r_i)}-\Delta_\lambda^{(-)})^2} 
\sin^2\left[\frac{\Delta_m^{(r_i)}-\Delta_\lambda^{(-)}}{2}z\right] 
+\frac{1}{(\Delta_m^{(r_i)}-\Delta_\lambda^{(+)})^2} 
\sin^2\left[\frac{\Delta_m^{(r_i)}-\Delta_\lambda^{(+)}}{2}z\right] 
\right) 
\end{equation} 
where we have kept only the resonant term, which depends 
on the sign of $\Delta_m^{(r_i)}-\Delta_\lambda$, and 
where we have defined 
\begin{equation} 
\Delta_\lambda^{(\pm)}\equiv\Delta_\lambda\pm\Delta_0. 
\end{equation} 
Now if $|\Delta_m^{(r_i)}-\Delta_\lambda^{(+)}|\ll|\Delta_M|$ or 
$|\Delta_m^{(r_i)}-\Delta_\lambda^{(-)}|\ll|\Delta_M|$ we find a strong mixing 
regime, meaning that because of the resonance there will exist a mode 
for which the probability is enhanced. 
 
In conclusion, the important scale that fixes the photon effective mass 
is now $\Delta_\lambda^{(\pm)}\sim\Delta_0$ if we are in a regime where  
$|\Delta_\lambda|\ll\Delta_0$. We can then have the 
same discussion as in the previous sections but with $\beta$ defined 
as 
\begin{equation} 
\beta=\frac{\Delta_m}{\Delta_\lambda^{(\pm)}}, 
\end{equation} 
according to the sign of $\Delta_\lambda$. The length scale 
$\lambda_\gamma$ is now given by $(\omega\Delta_0)^{-1/2}$ and 
it follows that we expect  
the two following effects: 
\begin{enumerate} 
\item by increasing $\Delta_0$ we can hope to make $\beta$ as small as 
wanted and thus to get a large enhancement of the oscillation 
probability. What happens is that the scale $\Delta_\lambda$ is replaced 
by $\Delta_0$ and thus that a departure from the four dimensional case 
will be observed if $(\omega\Delta_0)^{-1/2}<R$. When the field is 
homogeneous, the scale $\Delta_\lambda^{-1}$ is usually very large 
compared to $R$ (see \S~\ref{par6} and \S~\ref{par7}), which implies 
that there is little hope to see any effect of the extra--dimensions. By 
using an inhomogeneous field, we change the scale associated with the 
photon effective mass wich is now gouverned by $\Delta_0^{-1}$ that can 
be tried to be lowered to a scale close to $R$. 
\item whatever the sign of $\Delta_\lambda$, we expect to have 
strong mixing occuring for all values of $\Delta_0$ such that 
$$|\Delta_m^{(r_i)}-(\Delta_\lambda\pm\Delta_0)|\ll\Delta_M.$$ 
By varying slowly $\Delta_0$ or $\omega$, we expect to see a series of strong 
and weak mixing regimes. 
\end{enumerate} 
The amplitude of these two effects will be discussed in the last section 
of this article. 
  
\section{Application to astrophysics and cosmology}\label{par6}

Magnetic fields are observed in most astrophysical systems but the 
origin of galactic and cosmological magnetic fields  is still unknown \cite{kronberg94}. 
A possibility is that these 
fields have a primordial origin since such a magnetic 
field can be generated in a number of early universe mechanisms 
\cite{enqvist98} such as in collisions of bubbles produced in a first 
order phase transition \cite{kibble95} or during an inflationary phase 
\cite{turner88}. 
  
The efficiency of the photon--graviton and of the photon--axion mixing  
depends both on the value of the magnetic field and on the spatial  
extension of this field, $\Lambda_c$ say. We study the order of  
magnitude of these mixings on the cosmic microwave background, on  
pulsars and magnetars. 
  
The required quantities for our discussion are  
$\Delta_M$, $\Delta_m$,  $\Delta_{\rm plasma}$ and  
$\Delta_{\rm QED}$ respectively given by equations (\ref{deltas}) and  
(\ref{deltabis}). It is usefull to rewrite these quantities numerically as  
\begin{eqnarray}  
  \frac{\Delta_M}{1\,{\rm cm}^{-1}}&=&  
       4\times10^{-25}\left(\frac{H_0}{1\,\rm G}\right) 
\qquad\hbox{(graviton)},  
       \nonumber\\  
  \frac{\Delta_M}{1\,{\rm cm}^{-1}}&=&  
       2\times10^{-16}\left(\frac{H_0}{1\,\rm G}\right)  
       \left(\frac{f_{\rm PQ}}{10^{10}\,\rm GeV}\right)^{-1} 
       \quad \hbox{(axion)},  
       \nonumber\\  
  \frac{\Delta_m}{1\,{\rm cm}^{-1}}&=&  
       \frac{-2.5\times10^{28}}{\left(2.5\times10^{15}\right)^{4/n}}  
  \left(\frac{M_D}{1\,\rm TeV}\right)^{2+4/n}\left(\frac{\omega}{1\,\rm  
  eV}\right)^{-1},\nonumber\\  
  \frac{\Delta_{\rm plasma}}{1\,{\rm cm}^{-1}}&=&  
          -3.6\times10^{-17}\left(\frac{\omega}{1\,\rm eV}  
  \right)^{-1}\left(\frac{n_e}{1\,{\rm cm}^{-3}}\right),\nonumber\\  
  \frac{\Delta_{\rm QED}}{1\,{\rm cm}^{-1}}&=&  
     1.33\times10^{-27}\left(\frac{\omega}{1\,\rm eV}\right)  
  \left(\frac{H_0}{1\,\rm G}\right)^2,  
\label{deltanum}  
\end{eqnarray}  
where we have used the facts that $1\,{\rm eV}\simeq 5\times  
10^{4}\,{\rm cm}^{-1}$, $1\,{\rm G}\simeq 1.95\times10^{-2}\,{\rm  
eV}^2$ in the natural Lorentz-Heaviside units where  
$\alpha=e^2/4\pi=1/137$ and the expression of the extra-dimensions 
radius 
\begin{equation}\label{Rnum} 
R=\left(2.5\times10^{15}\right)^{2/n}10^{-12}\left(\frac{M_D}{1\,\rm 
Tev}\right)^{-1-2/n}\,{\rm eV}^{-1}. 
\end{equation} 
We now restrict to the case $n=2$.

\subsection{Cosmic microwave background}\label{par6_1}  
  
It has been shown that the isotropy of the cosmic microwave background  
(CMB) puts a limit on the  
present value of a spatially homogeneous magnetic field to $(B_0/1  
{\rm G})\leq 6.8\times10^{-9}(\Omega_0 h^2)^{1/2}$  
\cite{barrow97,peter}. A comparable bound has also been obtained for  
spatially inhomogeneous magnetic fields \cite{subramanian98}.  
We study the magnitude of the photon--graviton conversion on the  
two following examples:  
\begin{itemize}  
\item {\it Large scales}: we assume that we have a homogeneous  
magnetic field on the scale of the Hubble radius with  
\begin{equation}  
  {H_0}\simeq 6\times10^{-9}\, \rm G.  
\end{equation}  
The CMB photons are observed as a black body with a temperature of 2.7~K  
so that we approximatively have photons of energy \cite{smoot1,mather}  
\begin{equation}  
     \omega\simeq 10^{-5}-10^{-3}\,\rm eV.  
\end{equation}  
The caracteristic size of the system is the size of the Hubble radius  
\begin{equation}  
\Lambda_c=3000 h^{-1}\, {\rm Mpc}\simeq 10^{28}\,{\rm cm}  
\end{equation}  
where $h$ is the reduced Hubble parameter. We also estimate the electronic  
density today to be about (see e.g. \cite{peebles93}) 
\begin{equation}  
n_e\simeq10^{-7}\,{\rm cm}^{-3}.  
\end{equation}

\item {\it Degree scales}: we assume a homogeneous magnetic field  
on the size of the Hubble radius at the last scattering surface. Since  
the magnetic field scales like (scale factor)$^2$ and the energy of   
the photon as (scale factor)$^{-1}$, we assume  
a magnetic field of  
\begin{equation}  
  H_0\simeq 6\times10^{-3} \,\rm G 
\end{equation}  
and consider photons of energy  
\begin{equation}  
  \omega\simeq 10^{-2}-1 \,\rm eV 
\end{equation}  
at a redshift of $z\simeq1000$.  
The characteric size of the system is given by the Hubble radius at  
decoupling, i.e.  
\begin{equation}  
\Lambda_c=3\times10^{23}h^{-1}\,{\rm cm},  
\end{equation}  
and the electronic density at the time of decoupling  
is of order (see e.g. \cite{peebles93})  
\begin{equation}  
n_e\simeq  10^{-3}{\rm cm}^{-3}.  
\end{equation}  
\end{itemize}  
  
\begin{figure}  
\centering  
\epsfig{figure=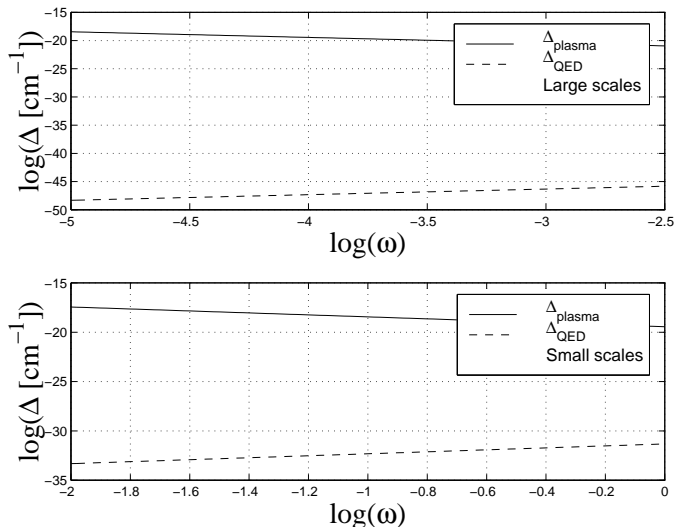,width=9cm}  
\caption{$\Delta_{\rm QED}$ (dash line) and $|\Delta_{\rm plasma}|$   
(solid line) for the CMB on large (up) and small (down) angular scales.  
We see that we always have $\Delta_{\rm QED}\ll\Delta_{\rm plasma}$. }  
\label{CMB2}  
\end{figure}

The main idea is that, since photons are converted into either 
gravitons or axions, some anisotropies must be induced on the scale of 
homogeneity of the magnetic field, mainly because of the angular 
dependence of the conversion rate. The effect between a direction 
parallel and direction perpendicular to the magnetic field must not 
exceed the observed CMB temperature anisotropy.  The anisotropy of the 
CMB temperature between the directions perpendicular to the magnetic 
field (where the effect of mixing is maximum) and parallel to it 
(where there is no mixing effect) is then of order 
\begin{equation}  
\frac{\Delta T}{T}\simeq\left.\frac{\Delta T}{T}\right|_\perp-  
\left.\frac{\Delta T}{T}\right|_{||}\simeq P(\gamma\rightarrow g).  
\end{equation}  
Observationally, we have the constraint \cite{smoot1} that  
\begin{equation}  
\frac{\Delta T}{T}< 10^{-5}.  
\end{equation}  
From figure \ref{CMB2}, we deduce that  
in both cases, $|\Delta_{\rm QED}|\ll|\Delta_{\rm plasma}|$ so that 
$\Delta_\lambda\simeq\Delta_{\rm plasma}$ and thus $\beta>0$. 
In the two considered regimes we have: 
 
\begin{enumerate}  
\item {\it Large angular scale}:  
\begin{eqnarray} 
\alpha&\in&6.6\times\left[10^{-15},10^{-13}\right]\nonumber\\ 
\beta&\simeq&1.1\times10^{21}\left(\frac{M_D}{1\,\rm TeV}\right)^4. 
\end{eqnarray} 
Thus, we are always in a regime where $\alpha\ll1$, $\beta>0$ and 
$|\beta|\gg1$ thus we expect at most effects of order $\alpha^2$ which 
are completely unobservable. 
 
\item {\it Small angular scale}: 
\begin{eqnarray} 
\alpha&\in&6.6\times\left[10^{-10},10^{-8}\right]\nonumber\\ 
\beta&\simeq&1.1\times10^{17}\left(\frac{M_D}{1\,\rm TeV}\right)^4. 
\end{eqnarray} 
We are always in a regime where $\alpha^2\ll1$, $\beta>0$ 
and $|\beta|\gg1$ and, as in the previous case, there 
will be no observable effect.  
\end{enumerate} 
 
From this results, with see that $\beta$ is always too large to have any
enhancement of the probability. Moreover in both cases the oscillation length, $\ell_{\rm 
osc}$, with the lightest KK mode (as well as the oscillation length with any massive KK mode) is much 
smaller than $\Lambda_c$, and the mixing angle with the graviton zero mode is very small. The effects are the same as in a standard four dimensional 
spacetime and thus negligible \cite{chen95,cillis96}. Note that, in theory, we should have included the 
expansion of the universe but this will not change the result 
drastically. A detailed study of the photon--graviton mixing in an 
expanding four dimensional spacetime can be found in \cite{magueijo94} 
and a discussion of the effects of the inhomogeneity of the field in 
\cite{cillis96}.

\subsection{Pulsars}\label{par6_2}  
  
As proposed by many authors (see e.g.  
\cite{raffelt88,morris,iwamoto84}), axions could be produced in  
the interior of neutron stars in nucleon--nucleon collisions. This  
would constitute the main cooling mechanism for these stars and  
thus puts limit on the axion production flux and mass. Such a  
production of KK gravitons in higher dimensional theories also  
exist and can be used to put bounds on the mass scale $M_D$  
\cite{barger99,cullen99}.  
   
As originally proposed by Morris \cite{morris} (see also 
\cite{raffelt88}), this axion (and now the KK gravitons) flux may be 
detectable by the secondary photons produced through the mixing with 
these particles in the neutron star magnetosphere magnetic field. These 
photons have a typical energy of 
\begin{equation}  
 \omega\simeq 10^4\,\rm eV,  
\end{equation}  
i.e. of order of the average value of the neutron star interior  
temperature (about 50~keV). The primary photons can be well  
approximated by a black body spectrum with a temperature of  
$T_{NS}\simeq1\,{\rm keV}$ typical for the surface temperature of  
such stars.  
   
The idea is to detect a distortion of the star spectrum due both to  
the secondary photons and to the oscillation of primary photons. 
 The typical value of the magnetic field in the  
neutron star magnetosphere is  
\begin{equation}  
 H_0\simeq10^{12}\,\rm G  
\end{equation}  
on a characteristic size of  
the system is of order of the neutron star size  
\begin{equation}  
\Lambda_c\simeq10\,{\rm km}.  
\end{equation}  
Indeed, one cannot neglect the effect of the magnetospheric  
plasma and we estimate its density \cite{shapiro83,goldreich69} as  
\begin{equation}\label{nenum}  
n_e\simeq7\times10^{-2}\left(\frac{H_0}{1\,{\rm G}}\right)  
\left(\frac{P}{1\,{\rm s}}\right)^{-1}  
\end{equation}  
where $P$ is the period of the pulsar and will be   
assumed to be about 1 second in the following.  
 
\begin{figure}  
\centering  
\epsfig{figure=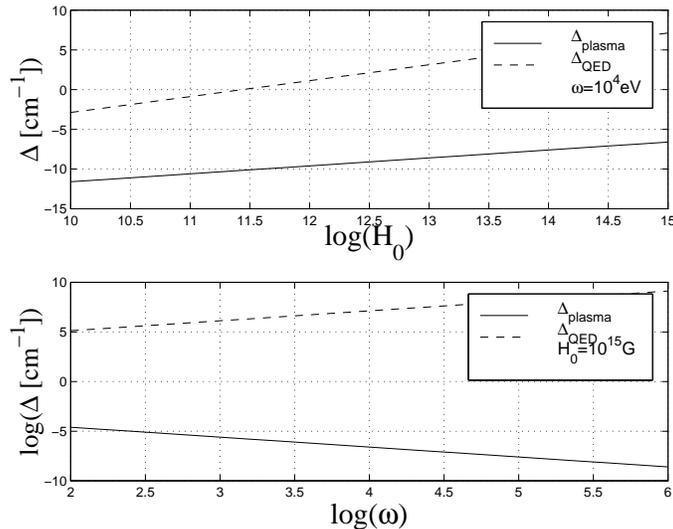,width=9cm}  
\caption{Variation of $\Delta_{\rm plasma}$ (solid line) and   
$\Delta_{\rm QED}$ (dash line) respectively in function of the magnetic field  
$H_0$ for a photon of $\omega=10^4$~eV(top) and in terms of the  
frequency for a field of $H_0=10^{15}$~G (bottom).}   
\label{puls3}  
\end{figure}

According to figure \ref{puls3}, we deduce that $\Delta_{\rm QED} 
\gg|\Delta_{\rm plasma}|$ so that $\beta<0$ and $\Delta_\lambda\simeq 
\Delta_{\rm QED}$. Then, it follows 
\begin{eqnarray}\label{173} 
\alpha&\simeq&3\times10^{-14}\nonumber\\ 
|\beta|&\simeq&3\times10^{-8}\left(\frac{M_D}{1\,\rm TeV}\right)^4 
\end{eqnarray} 
and we are always in a regime where $\alpha\ll1$, $\beta<0$ and
$|\beta|\leq1$ and where the characteristic size of the system is far
larger than the oscillation length.  We expect $M_D$ to be of order
$1-100$~TeV, so that we will get an amplification of order $1-10^7$ but
still unobservable.  From (\ref{173}) we see that, contrary to the
microwave background, the dominant length scale of the system is
$\Delta_{\rm QED}^{-1}$ so that
\begin{equation}\label{lr} 
\frac{\lambda_\gamma}{R} 
\simeq2.4\times10^{12}\left(\frac{\omega}{1\,\rm eV}\right)^{-1} 
\left(\frac{H_0}{1\,\rm G}\right)^{-1}\left(\frac{M_D}{1\,\rm TeV}\right)^{2} 
\end{equation} 
which is smaller than unity for the typical value of magnetic 
field and wavelength considered here. 
By going to higher frequencies and higher magnetic fields we 
may get a larger amplification and thus a  larger effect 
of the extra--dimensions, this is mainly the reason why we 
will turn to magnetars in the following paragraph.  
 
To finish, let us note however that in very strong magnetic fields one 
must take into account the photon splitting \cite{adler71,adler2} which 
will compete with the photon--graviton mixing. We do not discuss this 
effect here. 
 
\subsection{Magnetars and Gamma--Ray Bursts}  
\label{par5_3}  
  
Magnetars are pulsars with superstrong magnetic field such as SGR  
1806-20 \cite{kouveliotou,kouveliotou2} where $H_0\simeq 8\times10^{14}\,{\rm  
G}$, i.e. two orders of magnitude higher than for ordinary radio  
pulsars. This object is associated with soft gamma ray bursts of 
energy of order $1\,{\rm keV}-100 {\rm keV}$. Other examples are GB790305 
\cite{paczynski92} and IE1841-045 \cite{vasisht97} and such 
observations are supported by models where the gamma ray bursts 
are triggered by cracking of the neutron star crust due to the 
magnetic stress \cite{thompson95,thompson96}.  
 
\begin{figure} 
\centering 
\epsfig{figure=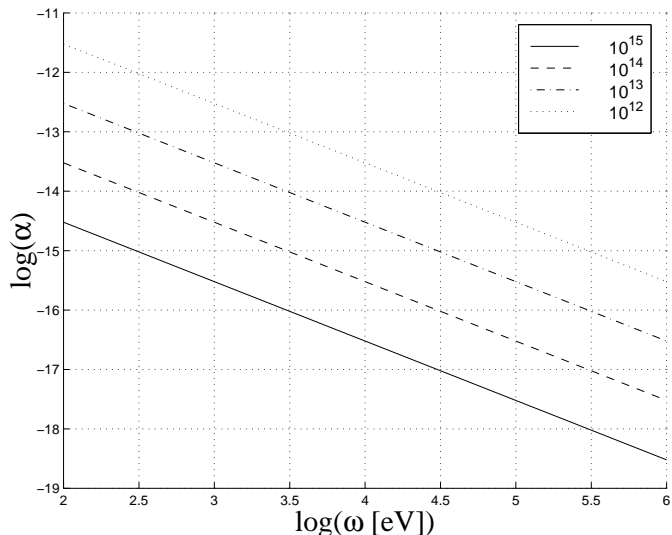,width=9cm} 
\caption{Variation of $\alpha$  
with respect to the photon frequency for a magnetic field varying between
$H_0=10^{12}$~G and $H_0=10^{15}$~G.} 
\label{puls2} 
\end{figure}

So we consider a system such that 
\begin{equation} 
 H_0\simeq10^{12}-10^{15}\,{\rm G},\quad\Lambda_c\simeq10\,{\rm km}, 
\end{equation} 
and 
\begin{equation} 
  \omega\simeq 10^2-10^6 \, \rm eV
\end{equation}
Assuming that the electronic density
is well approximated by (\ref{nenum})\footnote{The pulsars
cited above have a period ranging from 4 to 10 seconds.}, we deduce that we
are in the regime $|\Delta_{\rm plasma}|\ll\Delta_{\rm QED}$
as long as
$$
\left(\frac{\omega}{\rm 1\, eV}\right)^2 \left(\frac{H_0}{\rm 1\,
G}\right) \gg 4\times 10^{9}$$ so that we can deduce that the QED
contribution always dominates in such object and then that
$\beta<0$. On figure \ref{puls2}, we depict the variation of $\alpha$
to show that we always have $\alpha\ll1$. 

Now, effects of the extra--dimensions will appear when
$\lambda_\gamma/R<1$ and this quantity varies typically from $10^{-18}$
to $10^{-25}$ for $n=2$ assuming $M_D\sim1\,\rm TeV$ (see equation
(\ref{lr})). Note also that $\omega^{-1}$ is of the order of $R$ and
that most of the effect comes from the fact that $\Delta_{\rm QED}$
becomes large compared with $R^{-1}$. Such objects may be interesting
to detect the effects of the extra--dimensions but more data and a
better understanding of soft gamma--ray bursts are needed before drawing
any conclusions.

\section{Laboratory experiments}\label{par7} 

Many experiments searching for light particles like axions were set up
(see e.g.
\cite{raffelt88,sikivie83,sikivie84,morris,bibber89,kim98,sikivie97} and
\cite{raffelt98} for a recent review).  We can classify the methods in the
two following categories:
\begin{itemize}
\item The {\it direct methods} in which a flux of axions coming from
some astrophysical source (Sun, supernovae...)  is tried to be converted
into photons through an external magnetic field. These transitions have
been used to put bounds on astrophysical axion fluxes and coupling
constant \cite{sikivie83,sikivie84}. One could think to use the same
kind of experiments to put constraints in the case of a mixing with a
large number of KK states. Let us first discuss the case of KK
gravitons.  The energy flux into KK gravitons from astrophysical object
cannot exceed the bound on the energy flux in axions in the usual four
dimensional case since otherwise the cooling rates of these objects will
be too high \cite{chang99}.  Let us then assume that the efficiency of
detection is maximum and that the experiment is designed to collect {\it
all} the emitted particles. Since each graviton is coupled with a much
lower coupling constant than the four dimensional axion coupling
constant accessible to these kinds of experiments, we do not expect that
such {\it direct} detection methods will be able to see any KK graviton
coming from astrophysical sources. In other words, since each KK
graviton is coupled at tree level only to the photon (and not to other
gravitons), there is no effect of the large number of KK states in
these experiments.

\item The {\it indirect methods} where one tries to detect the mixing of
the photon through its effect on a photon beam in a magnetic field, both
on its amplitude and polarisation. Now, the photon being coupled at tree
level to all KK states, one expects a departure from the
usual case. The effect on the polarisation of the beam comes from the
fact that for axions only the $\times$ component of electromagnetic wave
couples to the axions. For the gravitons, both polarisations evolve
according to the same equation but, due to the QED and Cotton--Mouton
birefringence, $\Delta_+\not=\Delta_\times$ which implies a phase shift
between them.
\end{itemize}

In the next paragraphs, we focuse on polarisation experiments to detect
the phase shift. We first compute this phase shift in a $D$ dimensional
spactime and discuss two kinds of experiments respectively in a static
and periodic magnetic field.  
We must emphasize here that the magnitude of the mixing with axions depends on the free parameter 
$f_{\rm PQ}$ (in contrast with the mixing with gravitons which magnitude is fixed by the value of the
Planck mass), so that these experiments may be able to put constraints on bulk axion models.

\subsection{Phase shift in $D$ dimensions}\label{par7_15}

Let us go back to the four dimensional case for axion. Then, from
(\ref{53}), we deduce that, starting from an initial state
$(A(0),G(0)=0)$, the two polarisations $+$ and $\times$ evolve
respectively as, omitting a global phase $\omega u$,
\begin{equation}\label{145}
A_+(u)=\hbox{e}^{-i\Delta_+u}A_+(0),
\qquad
A_\times(u)=\left(\hbox{e}^{-i\Delta_\times'u}\cos^2\vartheta+ 
            \hbox{e}^{-i\Delta_g'u}\sin^2\vartheta\right)A_\times(0).
\end{equation}
Now, restricting to the weak mixing regime where $\vartheta\ll1$, we can expand
the $\Delta'$ defined in (\ref{defDelta}) as
\begin{equation}\label{146}
\Delta_\times'\simeq\Delta_\times+\vartheta^2(\Delta_\times-\Delta_m),\qquad
\Delta_g'\simeq\Delta_m-\vartheta^2(\Delta_\times-\Delta_m).
\end{equation}
Expanding (\ref{145}) to second order and taking into account
(\ref{146}) leads to
\begin{equation}\label{147}
A_+(u)=\hbox{e}^{-i\Delta_+u}A_+(0),
\qquad
A_\times(u)=\left[1-i\vartheta^2\zeta+\vartheta^2(\hbox{e}^{i\zeta}-1)\right]
\hbox{e}^{-i\Delta_\times u}A_\times(0)
\end{equation}
with $\zeta\equiv(\Delta_\times-\Delta_m)u$. We deduce that the
two modes evolve relative to each other as
\begin{equation}
\frac{A_\times(u)}{A_+(u)}=\left[1-i\vartheta^2\zeta
+\vartheta^2(\hbox{e}^{i\zeta}-1)\right]\hbox{e}^{-i(\Delta_\times-\Delta_+)u}
\frac{A_\times(0)}{A_+(0)}.
\end{equation}
The relative phase and amplitude of $A_\times$ with respect to $A_+$ 
then evolve as
\begin{equation}
\left|\frac{A_\times}{A_+}\right|(u)\simeq \left[1-2\vartheta^2\sin^2
\left(\frac{\zeta}{2}\right)\right]\left|\frac{A_\times}{A_+}\right|(0),\qquad
\phi(u)\simeq \left[(\Delta_+-\Delta_\times)u-\vartheta^2(\zeta-\sin\zeta)\right]
\phi(0).
\end{equation}
When we neglect the mixing effect (i.e. $\vartheta=0$), there is a phase shift
due to the QED and Cotton--Mouton birefringence. The extra phase shift due to
the fact that only one polarisation of the axion is affected by the mixing 
has been used to design experiments to put constraints on the axion parameters 
(see e.g. \cite{raffelt88,sikivie84,bibber89,maiani86}). 

In the case of a graviton, the two polarisations are mixed in 
the same way, so that the same computation leads to
\begin{eqnarray}
\left|\frac{A_\times}{A_+}\right|(u)&\simeq&\left(1-2\left[\vartheta^2_\times
\sin^2\left(\frac{\zeta_\times}{2}\right)-\vartheta^2_+
\sin^2\left(\frac{\zeta_+}{2}\right)\right]\right)
\left|\frac{A_\times}{A_+}\right|(0),\nonumber\\
\phi(u)&\simeq&\left[(\Delta_+-\Delta_\times)u
+\vartheta^2_+\left(\zeta_+-\sin\zeta_+\right)-
\vartheta^2_\times\left(\zeta_\times-\sin\zeta_\times\right)
\right]\phi(0)
\end{eqnarray}
(we now have to keep the index $\lambda$ on $\vartheta$ and on $\zeta$).
In four dimension, $\Delta_m=0$ so that the phase shift depends only on
the QED and Cotton--Mouton parameters. Its amplitude is proportional to
$\Delta_M$ so that it is roughtly 10 orders of magnitude
lower than for axions.\\

Let us now compute the phase shift in a $D$ dimensional spacetime.
We assume that we are in the weak mixing regime and apply the method of
\S~\ref{par4}. As seen on (\ref{147}), we must compute the solution of
(\ref{heisen1}) up to second order.  Following the same lines as for the
computation leading to (\ref{V1}), we can show that $\vec{\cal V}_{\rm
int}^{(2)}$ is explicitely given by
\begin{equation}\label{V2}
\vec{\cal V}_{\rm int}^{(2)}(u)=-\sum_q\int_0^udy\,\Delta_M(y)
\int_0^ydz\,\Delta_M(z)\,\hbox{e}^{i\int_y^z [\Delta_m^{(q)}-
\Delta_\lambda(x)]dx} A(0)\vec{\cal A},
\end{equation}
From which we deduce that, starting from a pure photon state,
the polarisation $\lambda$ of the photon evolves as
\begin{equation}\label{V3}
A_\lambda(u)=\left[1-\sum_q\int_0^udy\,\Delta_M(y)
\int_0^ydz\,\Delta_M(z)\,\hbox{e}^{i\int_y^z [\Delta_m^{(q)}-
\Delta_\lambda(x)]dx}\right] A_\lambda(0).
\end{equation}
In the case of a homogeneous field, we can extract from
(\ref{V3}) the relative phase of the polarisation $\times$
with respect to the polarisation $+$ for the case of
KK gravitons and bulk axions. In the latter case, only
the polarisation $\times$  evolves according to (\ref{V3}) whereas
the polarisation $+$ evolves according to (\ref{147}) so that
\begin{eqnarray}
\phi(u)&=&(\Delta_+-\Delta_\times)u-\alpha^2\sum_{i\geq1}\frac{s_i}{
(1-\beta^{(\times)}_i)^2}\left[(\Delta_\times-\Delta_m^{(r_i)})u-
\sin(\Delta_\times-\Delta_m^{(r_i)})u\right]\phi(0)\qquad \hbox{(axion)}
\label{P1}\\
       &=&(\Delta_+-\Delta_\times)u-\sum_{i\geq1}s_i
\left\lbrace\frac{\left[(\Delta_\times-\Delta_m^{(r_i)})u-
\sin(\Delta_\times-\Delta_m^{(r_i)})u\right]}{(1-\beta^{(\times)}_i)^2}
\alpha^2_\times
\right.\nonumber\\
&&\left.\qquad\qquad\qquad\qquad\qquad\qquad
\qquad-\frac{\left[(\Delta_+-\Delta_m^{(r_i)})u-
\sin(\Delta_+-\Delta_m^{(r_i)})u\right]}{(1-\beta^{(+)}_i)^2}\alpha^2_+
\right\rbrace\qquad\hbox{(graviton)}
\label{P2}
\end{eqnarray}
with the notations used before. We split this result in three parts
as
\begin{equation}
\phi=\phi_{\rm QED}+\phi_{\rm CM}+\phi_{KK}
\end{equation}
where the two first terms are the phase shifts due to vacuum polarisation
and the Cotton--Mouton effect and are obtained by setting $\alpha=0$ in
(\ref{P1}-\ref{P2}). The third term is the specific phase shift
associated with the mixing of the photon with either bulk axions
or Kaluza--Klein gravitons.

\subsection{Polarisation experiments}\label{par7_2}

We now discuss two kinds of experiments designed to detect the
mixing induced phase shift. As typical parameters we take 
$\omega\simeq2\,{\rm eV}$ for the laser beam and a magnetic field
which might be as strong as $H_0=10^5$~G. With these values, we have
(for $n=2$)
\begin{eqnarray}
&&\frac{\Delta_M}{1\,{\rm cm}^{-1}}\simeq4\times10^{-20},\quad
\hbox{(graviton)}\qquad
\frac{\Delta_M}{1\,{\rm cm}^{-1}}\simeq2\times10^{-11}\left(
\frac{f_{\rm PQ}}{10^{10}\,{\rm GeV}}\right)^{-1}
,\quad\hbox{(axion)}\nonumber\\
&&
\frac{\Delta_{\rm QED}}{1\,{\rm cm}^{-1}}\simeq3\times10^{-17},\quad
\frac{\Delta_{\rm plasma}}{1\,{\rm cm}^{-1}}\simeq-1.3\times10^{-17}
\left(\frac{n_e}{1\,{\rm cm}^{-3}}\right),\quad
\frac{\Delta_m}{1\,{\rm
cm}^{-1}}\simeq-1.5\times10^{-4}\left(\frac{M_D}{1\,
\rm Tev}\right)^4.
\end{eqnarray}
Then, the QED and  plasma effects are of the same order of magnitude but,
since $\phi_{\rm QED}\gg\phi_{\rm CM}$, we neglect the Cotton-Mouton
effect and define the phase shift ratio as
\begin{equation}
{\cal R}_{KK}\equiv\frac{\phi_{KK}}{\phi_{\rm QED}}.
\end{equation}
One hopes
to be able to measure a ${\cal R}_{KK}$ of order 0.1 \cite{raffelt88}.

\subsubsection{Multiple path experiments}

The idea is to make a laser beam reflect between two mirrors distant of $l$.
Since the mirror are transparent to the axions and gravitons,
the phase shift after $N$ paths will be $N\phi(l)$ so that,
in the case of axions, ${\cal R}_{KK}$ is given by
\begin{equation}
{\cal R}_{KK}=\left(\frac{\Delta_M}{\Delta_+-\Delta_\times}\right)
\sum_{i\geq1}s_i\frac{\Delta_M}{\Delta_\times-\Delta_m^{(r_i)}}
\left[1-\frac{\sin(\Delta_\times-\Delta_m^{(r_i)})l}{(\Delta_\times
-\Delta_m^{(r_i)})l}\right].
\end{equation}
The decrease of the amplitude of the photon beams due to the
creation of axions is
\begin{equation}
\delta I \simeq 4\alpha^2N\sum_{i\geq1}\frac{s_i}{(1-\beta_i)^2}
           \sin^2\left[\frac{\Delta_\times-\Delta_m^{(r_i)}}{2}l\right].
\end{equation}
Then, an enhancement of both the phase shift and the variation
of the beam are expected due to the sums over all states. With
the experimental values specified above, we have 
\begin{equation}
\frac{\lambda_\gamma}{R}\simeq 10^8
\left(\frac{M_D}{1\,\rm TeV}\right)^2,
\end{equation}
which is larger than unity. Then, we think that experiments
in homogeneous magnetic fields will not probe the
extra--dimensions since it will require to work with
very high magnetic fields. Note that when we span the
electromagnetic spectrum from the infrared to the X--ray, 
$\lambda_\gamma/R$ varies in the range
\begin{equation}
\frac{\lambda_\gamma}{R}\simeq 4\times(10^8-10^2)
\left(\frac{M_D}{1\,\rm TeV}\right)^2.
\end{equation}

\subsubsection{Effect of a periodic field}

As seen in \S~\ref{7b}, one can hope to enhance the mixing effect by
using a periodic magnetic field. For that purpose we need the pulsation
$\Delta_0$ to dominate over $\Delta_\lambda$ and $\beta$ to be small
compared to unity. With the previous numerical values, the first
condition rewrites as $\Delta_0 >10^{-17}\,{\rm cm}^{-1}$ and will be
satisfied easily.  Using (\ref{Rnum}), the second condition gives for a
six dimensional spacetime
\begin{equation}\label{195}
\Delta_0^{-1}<2.5\times10^2\left(\frac{M_D}{1\,\rm TeV}\right)^{-4}
\left(\frac{\omega}{1\,\rm eV}\right)\, {\rm cm}.
\end{equation}
As stressed before, we have a departure from the four dimensional
behaviour only if $\lambda_\gamma<R$. Now, since the magnetic field
varies on a scale $\Delta_0^{-1}$, $\lambda_\gamma$ is gouverned by
$\Delta_0$ instead of $\Delta_\lambda$.
Then an effect will appear only if we manage to create a field
that can vary on scales of the order of the centimeter.

In \S~\ref{7b} we also quoted the possibility of having a series
of strong mixing regimes, which is specific of the existence
of extra--dimensions. This requires that 
$|\Delta_m^{(r_i)}-(\Delta_\lambda\pm\Delta_0)|\ll\Delta_M\sqrt{s_i}$
and can be performed either by varying $\Delta_0$ or $\omega$.
Let us assume that $\Delta_0$ is fixed, since $\Delta_0\gg\Delta_M$
we have a strong mixing regime for the pulsations
defined by
\begin{equation}
\omega^{(\vec p)}=\frac{\vec p^2}{2R^2\Delta_0}
\end{equation}
with a width of
\begin{equation}
\delta\omega^{(\vec p)}=\frac{\vec p^2}{2R^2\Delta_0}
\frac{\Delta_M}{\Delta_0},
\end{equation}
that is
\begin{equation}
\frac{\omega^{(\vec p)}}{1\,\rm eV}=2\times10^{-4}\vec p^2
\left(\frac{M_D}{1\,\rm TeV}\right)^2\frac{\Delta_0^{-1}}{R},\qquad
\frac{\delta\omega^{(\vec p)}}{1\,\rm eV}=10^{-3}\vec p^2
\left(\frac{\Delta_0^{-1}}{R}\right)^2
\left(\frac{\Delta_M}{1\,\rm cm^{-1}}\right).
\end{equation}
For instance if we assume
that $R$ is of order of the millimeter and that we consider
a field varying on the order of the meter, we get for axions that
\begin{equation}
\frac{\omega^{(\vec p)}}{1\,\rm eV}=2\times10^{-2}\vec p^2
\left(\frac{M_D}{1\,\rm TeV}\right)^2,\qquad
\frac{\delta\omega^{(\vec p)}}{1\,\rm eV}=2\times10^{-10}\vec p^2.
\end{equation}

\section{Conclusion} 
 
We have used the property that, in an external magnetic field, the KK
gravitons and axions couple at tree level to photons to show that there
exists a mixing between the photon and these particles. We have computed
this mixing and compared our result to the photon-graviton and
photon-axion mixings in four dimensions.  The main difference comes from
the fact that the mixing matrix is now infinite and that a photon couples
to a large number of massive particles.  We then have discussed the
physical implications of these phenomena in a general $D$ dimensional
universe.  This leads us to conclude that for most astrophysical objects
the effect of photon-KK gravitons mixing will be unobservable.\\

The main points of this study are: 
\begin{itemize} 
  \item We describe how to deal with the mixing between the
photon and a large number of light particles. This extends
the former results to the case of $D$ dimensional universes
where a photon can mix with all the Kaluza--Klein gravitons
and possibly  with bulk axions. In (\ref{ppp}--\ref{y}), we
gave the exact expression for the oscillation probability
and we then discussed its amplitudes first qualitatively and then in
a five and in a six dimensional spacetime.
  \item When $\lambda_\gamma<R$, i.e. if
the caracteristic length scale associated with the photon 
wavelength and effective mass is smaller than the radius of
the extra--dimensions, there is a departure from the four dimensional
effect. Otherwise, the first KK mode is too heavy to
be excited and everything, in general, reduces to the four dimensional
situation.
\item Two limiting regimes have been found: 
  \begin{itemize}
  \item A {\it large radius regime}  where the two following
  behaviours can appear
  \begin{itemize}
   \item In the {\it
  weak mixing} regime, we have shown that the oscillation probability
  can be obtained by summing over the individual oscillation
  probabilities and is then enhanced by a factor of order 
  $\beta^{-n/2}$ in comparison with the standard four dimensional case.
  In that case the solutions can be found either by solving the
  eigenvalues equation or by considering the equation of evolution as
  a Schr\"odinger equation and solving it iteratively in the
  interaction picture. The latter method generalises to the
  case of inhomogeneous magnetic fields but only applies if the
  oscillation probability is small compared to unity.
 \item In the {\it strong mixing} regime the photon mixes preferentially
  with a given KK modes, which is possible if the plasma effects dominate over
  the vacuum polarisation. In that case a complete transition is possible.
  We note that this effect is more likely to happen in a $D$ dimensional
  context than in a four dimensional spacetime and point out a specific
 effect of the extra--dimensions on the oscillation length.
 \end{itemize}
     \item A {\it small radius regime} where $R\ll\lambda_\gamma$ so that
  the spacing between to KK modes is very large compared to the
  photon characteristic length. In that case, the probability is
generally dominated by the contribution of the first state
 corresponding to the lightest particle 
 and we are back to the four dimensional case. A consequence of this is
 that we can have an observable signature of the extra--dimensions only if
 $\lambda_\gamma<R$. In most of the systems this cannot be achieved,
 mainly because $\Delta_\lambda$ is very small; the only favorable
 situation happens in strong magnetic fields such as in pulsars and magnetars
 and when we deal with a magnetic field which varies on
 a small enough typical scale.
 \end{itemize}
 \item We have shown that, in the case of graviton,
 the effect of this mixing although
enhanced is too small to be observed on the cosmic
microwave background and on astrophysical objects such as pulsars.
 However, we point out that the effects can be 
larger with bulk axions.
 \item We discussed laboratory experiments designed for the search of
axions in the light of this new framework and we computed the
phase shift (\ref{P1}--\ref{P2}) between the two polarisations 
of a photon entering a magnetic field. As for the probability, 
we show that the phase shift is enhanced by the existence of the KK
modes. In a periodic field, we show that the effect of the
extra--dimensions can be important and that there exists a series of
strong and weak mixing regimes that may be observed. More work to derive
bounds on $f_{\rm PQ}$ is however needed before drawing any conclusion.
 \item On a more technical level, we have shown how to 
diagonalise a general matrix of mixing. This kind of 
matrices appears in different situations in extra-dimensions physics 
and these results can be used in many problems and in 
particular for neutrino oscillations.
 \item To finish, let us stress some important comments. First,
we have assumed the validity of the Euler--Heisenberg Lagrangian to describe 
the vacuum polarisation. One has also to be aware that in strong magnetic
fields such as in pulsar magnetosphere, photon splitting 
\cite{adler71,adler2} will be in competition
 with the photon graviton oscillation. We have not compared 
the strength of these two effects but we expect the latter to be 
dominant in high magnetic fields, such as in the magnetosphere of
magnetars.  We have also concentrated on
gravitational waves even if there is also production of
scalar waves. They are thought to be negligible, at least in the
four dimensional case \cite{magueijo94}. Concerning the axions, the effects
may be more important than for gravitons, 
depending on the value of the coupling $f_{\rm PQ}$,
but we did not reconsider the bounds on the axion parameters. Further
work is needed in that direction. We also stress that in general one
needs a precise determination of their mass spectrum to compute the
coupling of each mode to the photon.
Most of these results were obtained for $n=1$ or
$n=2$ extra--dimensions for which the results respectively do not, or
only weakly, depend on the cut--off of the theory.
We also stress that depending on the exact physical situation the number
of KK modes with which the photon can oscillate coherently can be
drastically limited in comparison with the number of accessible KK
modes. These decoherence effect implies that the UV cut--off of the
theory $M_{\rm max}$ is expected to be much higher than the physical cut--off. A
consequence of this is that the results obtained in the cases $n=1$ and
$n=2$ can be extended to $n>2$ without depending on $M_{\rm max}$.
\end{itemize} 
 
\section*{Acknowledgments} 
We wish to thank l'\'Ecole de Physique Th\'eorique des Houches where
this work was initiated, P. Bin\'etruy, E. Dudas, J.F. Glicenstein,
R. Lehoucq, M. Lemoine, J. Mourad, O. Pene, P. Peter and J. Rich for
discussions.
 

\onecolumn 
\appendix 
\section{Diagonalisation of ${\cal M}$ in the general case}\label{appA} 
 
The goal of this appendix is to compute the eigenvalues and 
eigenvector of the matrix ${\cal M}$, to diagonalise it and to 
explain how to compute the probability of conversion of a photon 
into a graviton. 
 
We consider the $(N+2)\times(N+2)$ matrix defined by 
\begin{equation}\label{MND} 
  {\cal M}=\left(\begin{array}{cccccc} 
  \Delta_\lambda & \Delta_M &\cdots&\cdots&\cdots&\Delta_M\\ 
  \Delta_M&\Delta_m^{(0)}&0& \cdots&\cdots&0\\ 
  \vdots& 0 & \ddots&0&\cdots&0 \\ 
  \vdots&\vdots&0&\Delta_m^{(q)}&0 &\vdots\\ 
  \vdots&\vdots&\vdots&\ddots&\ddots&0\\ 
  \Delta_M&0&\cdots&\cdots&0&\Delta_m^{(N)} 
  \end{array}\right). 
\end{equation} 
We restrict to a finite matrix since there is a cut--off 
in the theory as discuseed in \S~\ref{par3_2} and all the notations
are defined in \S~\ref{par2}. 
 
Let us stress that the diagonalisation of this matrice is a purely 
technical point that appears often in extra-dimension physics (see 
e.g. \cite{dudas}). 

\subsection{Characteristic polynomial} 
 
These notations being fixed, we compute the characteristic polynomial 
${\cal P}(x)$ of the matrix ${\cal M}$ defined by 
\begin{equation}\label{polcar} 
  {\cal P}(x)\equiv\det\left({\cal M}-x I_{N+2}\right), 
\end{equation} 
where $I_{N+2}$ is the $(N+2)\times(N+2)$ identity matrix. 
Developping (\ref{polcar}) with respect to its first column leads 
to 
\begin{equation}\label{polcar2} 
  {\cal P}(x)=\left(\Delta_\lambda-x\right)\prod_{q=0}^N 
  \left(\Delta_m^{(q)}-x\right) 
  +\sum_{q=0}^N (-1)^{q+1}\Delta_M D_q(x), 
\end{equation} 
where $D_q$ is the determinant of the comatrix of the element $(q+2,1)$ 
given by 
\begin{equation}\label{comat} 
  D_q(x)=(-1)^q\Delta_M\prod_{\ell=0,\ell\not=q}^N 
  \left(\Delta_m^{(\ell)}-x\right). 
\end{equation} 
From (\ref{polcar2}) and (\ref{comat}), we deduce that the 
 characteristic polynomial is given by 
\begin{equation}\label{polcarfin} 
  {\cal P}(x)=\prod_{q=0}^N \left(\Delta_m^{(q)}-x\right) 
             \left[\Delta_\lambda-x-\Delta_M^2\sum_{q=0}^N 
             \frac{1}{\left(\Delta_m^{(q)}-x\right)} 
             \right]. 
\end{equation}

\subsection{Eigenvalues} 
 
The characteristic eigenvalue equation ${\cal P}(x)=0$ has $N+2$ real 
solutions since ${\cal M}$, being a symetric matrix, is 
diagonalisable. To find all his solutions, we rewrite 
(\ref{polcarfin}) as 
\begin{equation}\label{decpol} 
  {\cal P}(x)=A(x)B(x), 
\end{equation} 
with 
\begin{eqnarray} 
A(x)&=&\prod_{i=1}^{N_D}\left(\Delta_m^{(r_i)}-x\right)^{s_i-1}\nonumber\\ 
B(x)&=&\left(\Delta_\lambda-x\right)\prod_{i=1}^{N_D}\left(\Delta_m^{(r_i)} 
-x\right)-\Delta_M^2\sum_{i=1}^{N_D}\prod_{\ell=1,\ell\not=i}^{N_D}
s_\ell\left(\Delta_m^{(r_\ell)}-x\right). 
\end{eqnarray} 
We have two kind of eigenvalues: 
\begin{itemize} 
  \item $\Delta_m^{(r_i)}$: they are solutions of $A(x)=0$ and are of 
  order $s_i-1$ but are not solutions of $B(x)=0$ so that they are 
  solutions of ${\cal P}(x)=0$ with order $s_i-1$ and thus are 
  eigenvalues of ${\cal M}$ of the same order. 
 
  This gives us $\sum_{i=1}^{N_D}(s_i-1)=N+1-{N_D}$ eigenvalues of ${\cal M}$. 
 
  \item $x_i$: they are solutions of $B(x)=0$ and since $B(x)$ is a 
  polynomial of order ${N_D}+1$ and since ${\cal M}$ is diagonalisable, we 
  must have ${N_D}+1$ such eigenvalues. To find them, we rewrite $B(x)$ as 
  \begin{equation} 
  B(x)=\left[\prod_{i=1}^{N_D}\left(\Delta_m^{(r_i)}-x\right)\right] 
  \left[\Delta_\lambda-x-\Delta_M^2\sum_{j=1}^{N_D}\frac{s_j}{\left( 
  \Delta_m^{(r_i)}-x\right)}\right].  
  \end{equation}
  Let us stress
  that $x_i\not=\Delta_m^{(r_i)}$ since otherwise the cancellation 
  occuring in the first factor is offset by a divergence in the second 
  factor. It follows that the $x_i$ are solutions of 
  \begin{equation}\label{eqxi} 
  \Delta_\lambda-x=\Delta_M^2\sum_{i=1}^{N_D}\frac{s_i}{\left(
  \Delta_m^{(r_i)}-x\right)}. 
  \end{equation} 
This solution can be found numerically but we can find the main 
properties of these eigenvalues graphically 
 from which we deduce that (\ref{eqxi}) has ${N_D}+1$ 
{\it distinct} solutions that we order as 
\begin{equation} 
  \left(x_i\right)_{1\leq i\leq {N_D}+1}\qquad x_1<\ldots<x_{n+1}. 
\end{equation} 
\end{itemize} 
 
In conclusion, we have found the $N+2$ eigenvalues of ${\cal M}$ which 
split in $n$ eigenvalues $\Delta_m^{(r_i)}$ each with multiplicity 
$s_i-1$ and in ${N_D}+1$ distinct eigenvalues $x_i$. 
 
\subsection{Eigenvectors} 
 
To determine the eigenvectors $V$ solution of 
\begin{equation}\label{eqvec} 
  {\cal M}V=xV, 
\end{equation} 
we set 
\begin{equation}\label{vect} 
  V\equiv\left(v,u_0,\ldots,u_N\right) 
\end{equation} 
so that (\ref{eqvec}) reduces to the system 
\begin{eqnarray} 
&&\Delta_\lambda v+\Delta_M\left(\sum_{q=0}^Nu_q\right)=xv\\ 
&&\Delta_Mv+\Delta_m^{(q)}u_q=xu_q. 
\end{eqnarray} 
\begin{itemize} 
  \item If $x=\Delta_m^{(r_i)}$, the eigenvectors generate a 
  subspace of dimension $s_i-1$ a basis of which is given explicitely by 
\begin{equation}\label{vec1} 
V_{r_i+p}=\frac{1}{\sqrt{(p+1)(p+2)}}\left[-\sum_{\ell=0}^pG_{r_i+\ell} 
+(p+1)G_{r_i+p+1}\right],\qquad 0\leq p\leq s_i-2, 
\end{equation} 
where $\lbrace A,(G_q)_{0\leq q\leq N}\rbrace$ is the initial 
orthonormal basis where we have written ${\cal M}$ in (\ref{MND}). 
One can check that this is an orthonotmal family, i.e. that 
$$\langle V_{r_i+p}\mid V_{r_j+q} 
\rangle=\delta_{pq}\delta_{ij}.$$ 
  \item If $x=x_i$, for each eigenvalue we have a subspace of 
  dimension 1 generated by the unit vector 
\begin{equation}\label{vec2} 
V_{x_i}=\frac{1}{\sqrt{\Delta_M^{-2}+\sum_{q=0}^N\left(x_i-\Delta_m^{(q)} 
\right)^{-2}}}\left(\frac{1}{\Delta_M},\frac{1}{x_i-\Delta_m^{(0)}},\ldots, 
\frac{1}{x_i-\Delta_m^{(N)}}\right) 
\end{equation} 
in the basis $\lbrace A,(G_q)_{0\leq q\leq N}\rbrace$. It is 
easy to show that they satisfy 
$$\langle V_{x_i}\mid V_{x_j} \rangle=\delta_{ij},\qquad 
\langle V_{x_i}\mid V_{r_j+p} \rangle=0. 
$$ 
\end{itemize} 
We have given the explicit form of the $N+2$ eigenvectors of 
${\cal M}$. It is worthwile noting that the eigenstates 
$V_{r_i+p}$ mix the different KK modes together while letting the 
photon unaffected whereas the eigenstates $V_{x_i}$ mix the photon 
with the $N+1$ KK gravitons.

\section{Probability of oscillation in the general case}\label{appB} 
 
To compute the oscillation probability between a photon and 
gravitons in a constant magnetic field, we follow the method by 
Raffelt and Stodolsky \cite{raffelt88} and solve the equation of 
evolution (\ref{systfinal}) as 
\begin{equation} 
\vec{\cal V}(u)=\hbox{e}^{-i{\cal M}u}\hbox{e}^{-i\omega u} 
\vec{\cal V}(0),\label{solg} 
\end{equation} 
where $\vec{\cal V}\equiv\lbrace A,G^{(0)},\ldots,G^{(N)}\rbrace$. We 
decompose 
this vector on the eigenvectors basis as 
\begin{equation} 
\vec{\cal V}(0)=\sum_{i=1}^{N_D}\sum_{p=0}^{s_i-2}h_{i,p}(0)V_{r_i+p}+ 
\sum_{i=1}^{n+1}f_i(0) V_{x_i}, 
\end{equation} 
where $h_{i,p}(0)$ and $f_i(0)$ are $N+2$ coefficients. 
Injecting this decomposition in (\ref{solg}), we obtain 
\begin{equation} 
\vec{\cal V}(z)=\left[\sum_{i=1}^{N_D}\sum_{p=0}^{s_i-2}h_{i,p}(0) 
\hbox{e}^{-i\Delta_m^{(r_i)}u}V_{r_i+p}   + 
\sum_{i=1}^{{N_D}+1}f_i(0)\hbox{e}^{-ix_iu} 
V_{x_i}\right]\hbox{e}^{-i\omega u}. 
\end{equation} 
 
The probability of a photon to be converted in KK gravitons is 
obtained by considering the initial state $\vec{\cal 
V}(0)=\lbrace1,0,\ldots,0\rbrace$ 
and by computing 
\begin{equation} 
P(\gamma\rightarrow \gamma)=\left|\sum_{q=0}^N\langle 
G^{(q)}(z)\mid\vec{\cal V}(0) 
\rangle\right|^2. 
\end{equation} 
Since only the modes associated with the eigenvalues $x_i$ mix 
with the photon, we deduce that 
\begin{equation}\label{B5} 
P(\gamma\rightarrow g)=1-P(\gamma\rightarrow\gamma) 
=1-\left|\sum_{i=1}^{{N_D}+1}f^2_{x_i}\hbox{e}^{ix_iu}\right|^2, 
\end{equation} 
where the coefficients $f_{x_i}$ are 
\begin{equation}\label{B6}
f_{x_i}=\left[1+\sum_{q=0}^N\frac{\Delta_M^2}{\left( 
x_i-\Delta_m^{(q)}\right)^2}\right]^{-1/2}.
\end{equation}

\section{Upper bound on $|{\cal F}_J|$ and $|{\cal G}_J|$} 
\label{appD} 

The goal of this appendix is to give an upper bound on the absolute
value of the two functions ${\cal F}_{J}(y)$ and ${\cal G}_{J}(y)$ (defined in (\ref{deff}) and 
(\ref{C19}), (\ref{C20})) 
when $n=2$.
These majorations are used in section \ref{DD} to determine the solution
of the eigenvalues equation (\ref{yi1}) as well as the oscillation probability (\ref{ppp}) 
in a small coupling limit.
We also give an upper bound on $s_i$.
We first give a bound on 
\begin{eqnarray} 
{\cal F}_{i_1,i_2}(y)&\equiv&
 \alpha^2\sum_{i=i_1}^{i_2}\frac{s_i}{y-\beta_i}, \hbox{ and on}
  \label{fi1}\\ 
{\cal G}_{i_1,i_2}(y)&\equiv& \alpha^2\sum_{i=i_1}^{i_2}
  \frac{s_i}{\left(y-\beta_i\right)^2}.\label{gi1} 
\end{eqnarray} 
We recall that the $\beta_i$ 
are defined by $\beta_i\equiv\vec{p_i}^2 \beta\equiv p_i^2 \beta,$  
where  $\vec{p_i}$ is a pair $(n_i,m_i)$ of integers, and $p_i$ is defined by
the second equality. We assume all along this discussion that  
$\beta$ is positive and we order the $\beta_i$ as $\beta_i < \beta_{i+1}$. 
We stress that $\beta_1=0$. 

\subsection{$0 \leq y<\beta_{i_1} <\beta_{i_2}$} 

For $i\geq i_1$, each $\vec{p}_i^2$ belongs to a unique interval 
\begin{equation} 
  (p_{i_1}+k)^2\leq p_i<(p_{i_1}+k+1)^2, \label{interbeta} 
\end{equation} 
where $k$ is an integer. ${\cal M}_k$, the number of such $\vec{p}_i$, is
bounded by  four times the surface defined by
\begin{equation} 
  p_{i_1}+k \leq \sqrt{x_1^2+x_2^2} \leq p_{i_1}+k+1  
\end{equation}
in the real plan $(x_1,x_2)$, since we have at most four pairs for each
unit square cell. Thus, one has   
\begin{equation} 
  {\cal M}_k \leq 8 \pi (p_{i_1}+k+1).
\end{equation} 
One can then easily obtain a majoration in term of $(p_{i_1}+k)$:
\begin{equation} 
  {\cal M}_k \leq q (p_{i_1}+k),
\end{equation} 
with $q= 16 \pi$. On the other hand, 
for each $p_i$ satisfying (\ref{interbeta}), $|1 / (y-\beta_i)|$ is 
lower than $|1/(y-(p_{i_1}+k)^2\beta)|$, from which we
get the upper bound on ${\cal F}_{i_1,i_2}(y)$ 
\begin{equation} 
  \left|{\cal F}_{i_1,i_2}(y)\right| \leq \frac{\alpha^2 q}{\beta}
  \sum_{k=0}^{k_{\rm max}} {\sqrt{\beta}(p_{i1}+k) 
  \over \left(\sqrt{\beta}(p_{i1}+k)\right)^2-y } 
  \sqrt{\beta}. \label{bound2} 
\end{equation}  
$k_{\rm max}$ is defined such that
$p_{i_1}+k_{\rm max}<p_{i_2}<p_{i_1}+k_{\rm max}+1$. The r.h.s. of (\ref{bound2})
is nothing else but a Riemann sum associated with the function  
$f(x)= x / (x^2-y)$. Since $f(x)$ is decreasing for all $x^2>y$, 
we have 
\begin{eqnarray} 
  \left|{\cal F}_{i_1,i_2}(y)\right|\leq\frac{\alpha^2 q}{\beta}  
  \int_{\sqrt{\beta}(p_{i_1}-1)}^{\sqrt{\beta}(p_{i_1}+k_{\rm max})} 
  \frac{x dx}{x^2-y}
  \leq \frac{\alpha^2 q}{2 \beta}\ln\left({\beta(p_{i_1}+k_{\rm max})^2-y 
  \over\beta(p_{i_1}-1)^2-y}\right).\label{borne1} 
\end{eqnarray} 
Let us emphasize that (\ref{borne1}) assumes implicetly that 
$y<\beta (p_{i_1}-1)^2$. Otherwise, the contribution of the 
$\beta_{i}$ such that $\beta p_{i_1}^2\leq\beta_i<\beta(p_{i_1}+1)^2$
in the sum (\ref{fi1}) can be bounded by
\begin{equation} 
 {\alpha^2q\over\beta p_{i_1}^2-y}p_{i_1}.  
\end{equation} 
Since $\beta (p_{i_1}-1)^2 \leq y < \beta p_{i_1}^2$, we deduce that  
\begin{equation}
p_{i_1} \leq \left({\sqrt{y} \over \sqrt{\beta}} +1 \right),
\end{equation}
so that we get the majoration  
\begin{eqnarray}  \label{C12}
  |{\cal F}_{i_1,i_2}(y)|\leq{\alpha^2q\over
  (\beta p_{i_1}^2 -y)}\left({\sqrt{y}\over \sqrt{\beta}}+1\right) 
  +\frac{\alpha^2q}{\beta}\ln\left({\beta(p_{i_1}+k_{\rm max})^2-y 
  \over\beta p_{i_1}^2-y}\right). 
\end{eqnarray} 
With  similar arguments, on can show that, when $y<\beta
(p_{i_1}-1)^2$, $|{\cal G}_{i_1,i_2}(y)|$ is bounded by 
\begin{eqnarray} 
  |{\cal G}_{i_1,i_2}(y)|&\leq&\frac{\alpha^2 q}{\beta}  
  \int_{\sqrt{\beta}(p_{i_1}-1)}^{\sqrt{\beta}(p_{i_1}+k_{\rm max})} 
  \frac{x dx}{(x^2-y)^2}\leq\frac{\alpha^2 q}{2\beta} 
  \left(\frac{1}{\beta (p_{i_1}-1)^2 -y} -  
  {1\over \beta (p_{i_1}+k_{\rm max})^2-y} \right), 
\end{eqnarray} 
and that otherwise 
\begin{eqnarray} \label{C14}
  |{\cal G}_{i_1,i_2}(y)|&\leq&{\alpha^2q\over
  (\beta p_{i_1}^2-y)^2} \left({\sqrt{y}\over \sqrt{\beta}}+1\right) 
  +\frac{\alpha^2 q}{2 \beta} \left(\frac{1}{\beta p_{i_1} ^2 -y} 
  - {1 \over \beta (p_{i_1}+k_{\rm max})^2-y} \right). 
\end{eqnarray} 

The bounds (\ref{C12}) and (\ref{C14}) are valid also in the case where $y<\beta
(p_{i_1}-1)^2$.
\subsection{$0 \leq \beta_{i_1} <\beta_{i_2} < y $}

In that case, following the same line of reasoning, we obtain  
respectively for ${\cal F}_{i_1,i_2}$ and ${\cal G}_{i_1,i_2}$ and any $y$ satisfying the above condition
 
\begin{equation}  \label{C16}
  |{\cal F}_{i_1,i_2}(y)|\leq{\sqrt{y} \alpha^2 q \over \sqrt{\beta}
  ( y- \beta p_{i_2}^2 )}
  +\frac{\alpha^2 q}{2 \beta} \ln \left( {y- \beta (p_{i_2}-k_{\rm max})^2  
  \over y- \beta p_{i_2}^2}\right), 
\end{equation} 
\begin{equation}  \label{C18}
  |{\cal G}_{i_1,i_2}(y)|\leq {\sqrt{y} \alpha^2 q \over \sqrt{\beta}
  ( y- \beta p_{i_2}^2 )^2} 
  +\frac{\alpha^2 q}{\beta} \left(
  \frac{1}{y - \beta p_{i_2}^2} -{1 \over y- \beta (p_{i_2}-k_{\rm max})^2}  \right),
\end{equation} 

\subsection{bound on ${\cal F}_J$ and ${\cal G}_{J}$} 

We now  give a bound on the expression defined by 
\begin{eqnarray} 
  {\cal F}_{J}(y)&\equiv& \alpha^2\sum_{i=1,i \neq J}^{N_D} \label{C19}
  \frac{s_i}{y-\beta_i}, \\ 
  {\cal G}_{J}(y)&\equiv& \alpha^2\sum_{i=1, i \neq J}^{N_D} \label{C20}
  \frac{s_i}{\left(y-\beta_i\right)^2},  
\end{eqnarray} 
For $J$ defined by 
\begin{equation} 
\forall i \neq J,\qquad |y-\beta_i| \geq |y-\beta_J|.  
\end{equation}
One has thus $\forall i \neq J,\qquad |y-\beta_i| \geq \beta/2$, and in particular 
$\beta p_{J+1}^2 - y > \beta/2$ and $y-\beta p_{J-1}^2 > \beta/2$.
Using equations (\ref{C12}--\ref{C18})
as well as the value of $p_{\rm max}$ given\footnote{Note that the
energy cut--off $p_{\rm max}$ has to be regarded as a maximum one fixed
by the underlying quantum regularisation of the theory. However,
decoherence effects can reduce drastically the number of KK modes that
have to be taken into account. Thus, the bounds on ${\cal Q}$ and ${\cal
Q}'$ in a more realistic case are likely to be lower than the one given
here.} in (\ref{pmax2d}), one obtains 
easily the following bound on $|{\cal F}_J|$ and $|{\cal G}_J|$ 
\footnote{Similar bounds would also apply
to ${\cal F}(y)$ and ${\cal G}(y)$ when $y$ is so that 
$\forall i \in [1,N_D],\quad |y-\beta_i| \geq \beta / 2$.}. 
\begin{eqnarray}\label{C23}
|{\cal F}_{J}| &\leq&  {\cal Q} \frac{\alpha^2}{\beta^2} \hbox{ sup }
\left(  \sqrt{\beta y} , {\cal Q'} \right)  \label{C22}\\
|{\cal G}_J| &\leq&  {\cal Q} \frac{\alpha^2}{\beta^2} \hbox{ sup }
\left(\frac{\sqrt{y}}{\sqrt{\beta}},2 \right) \label{C24}
\end{eqnarray}
where  ${\cal Q}$ is a constant of order $10^3$ and ${\cal Q'}$ a constant of order 10. 
\subsection{bound on $s_{i}$}

In the real plane the euclidian distance between two different pairs of integers is bounded by $1$, so
that the number of pairs of integers on a given closed curve is always lower than the length of this curve.
It is then easy to obtain the following bound on $s_i$ which represents then number of different pairs of
integers on a circle of radius $p_i$.
\begin{equation}
s_i \leq 2 \pi p_i = 2 \pi \sqrt{{\beta_i \over \beta}}.  \label{C26}
\end{equation}
To finish, let us note that the properties of the series $s_i$ have been
studied by Gauss around 1800, see e.g.  [{\tt
http://mathworld.wolfram.com/rn.html}] for details and references.

\end{document}